\documentclass[12pt]{article}
\pdfoutput=1
\usepackage{tikz}
\usetikzlibrary{matrix,arrows,decorations.pathmorphing,decorations.pathreplacing,shadows,decorations.markings,patterns}
\usepackage{jheppub}
\usepackage{amsmath,amssymb,euscript,array,mathrsfs,appendix,ctable,marvosym}
\usepackage{arydshln}
\usepackage{todonotes}
\usepackage{graphicx}
\usepackage[normalem]{ulem}

\newcommand \arxivlink [1]{\href{http://arxiv.org/abs/#1}{\tt arXiv:#1}}

\DeclareMathAlphabet{\mathpzc}{OT1}{pzc}{m}{it}

\newcommand{\BH}[1]{\marginpar{\tiny BH: #1}}

\newcommand\blank[1]{#1}
\renewcommand\blank[1]{}
\def\Buildrel#1\over#2\under#3{\mathrel{\mathop{\kern0pt
#2}\limits^{#1}_{#3}}}

\def\LAX{{\mathfrak L}}
\def\PSU{\text{PSU}}
\def\Sp{\text{Sp}}
\def\CF{{\cal F}}

\def\PP{{\mathbb P}}

\def\mf{{\mathfrak f}}
\def\SO{\text{SO}}
\def\JJ{\mathscr{J}}

\def\msu{\mathfrak{su}}

\newcommand{\Tr}{\operatorname{tr}}
\newcommand{\STr}{\operatorname{str}}

\def\Bw{{\boldsymbol \omega}}

\def\Bdelta{{\boldsymbol\delta}}
\def\Bxi{{\boldsymbol\xi}}

\def\B0{{\boldsymbol 0}}
\def\BH{{\boldsymbol H}}

\def\Bp{{\boldsymbol p}}

\def\Bomega{{\boldsymbol\omega}}

\def\det{{\rm det}}
\def\SU{\text{SU}}

\newcommand{\Be}{\boldsymbol{e}}
\newcommand{\Balpha}{{\boldsymbol{\alpha}}}
\newcommand{\Bphi}{{\boldsymbol{\phi}}}

\def\Dbarslash{\,\,{\raise.15ex\hbox{/}\mkern-12mu {\bar D}}}
\def\Dslash{\,\,{\raise.15ex\hbox{/}\mkern-12mu D}}
\def\delslash{\,\,{\raise.15ex\hbox{/}\mkern-9mu \partial}}
\def\delbarslash{\,\,{\raise.15ex\hbox{/}\mkern-9mu {\bar\partial}}}
\def\Hslash{\,\,{\raise.15ex\hbox{/}\mkern-12mu H}}
\def\hatHslash{\,\,{\raise.15ex\hbox{/}\mkern-12mu {\widehat H}}}
\def\ptildeslash{\,\,{\raise-.08ex\hbox{/}\mkern-10mu {\tilde p}}}
\def\phatslash{\,\,{\raise-.08ex\hbox{/}\mkern-10mu {\hat p}}}
\def\Bpslash{\,\,{\raise-.08ex\hbox{/}\mkern-10mu {\boldsymbol p}}}

\newcommand{\MAT}[1]{\begin{pmatrix} #1\end{pmatrix}}

\newcommand{\EQ}[1]{\begin{equation}\begin{split} #1
\end{split}\end{equation}}

\title{Classical Spectral Curve of the AdS$\bf{_5{\boldsymbol\times}S^5}$ Lambda Superstring}
\author[a]{Timothy J. Hollowood,}
\author[b]{J. Luis Miramontes}
\author[a]{and Dafydd Price}
\affiliation[a]{Department of Physics, Swansea University, Swansea, SA2 8PP, U.K.}
\affiliation[b]{Departamento de F\'\i sica de Part\'\i culas and IGFAE,
Universidad de Santiago de Compostela, 15782 Santiago de Compostela, Spain}

\emailAdd{t.hollowood@swansea.ac.uk}
\emailAdd{jluis.miramontes@usc.es}
\emailAdd{dbprice@hotmail.com}

\abstract{The classical spectral curve for the worldsheet theory of the $\text{AdS}_5{\times} S^5$ lambda superstring is constructed. The lambda string is interpreted as a regularized, non-abelian T-dual of the $\text{AdS}_5{\times} S^5$ superstring with respect to the full $\text{PSU}(2,2|4)$ symmetry. The form of the curve is identified as the semi-classical limit of a set of Bethe ansatz equations for an XXZ type spin chain for the supergroup $\PSU(2,2|4)$ in contrast to the string in $\text{AdS}_5{\times} S^5$ which is XXX type.}

\notoc
\begin{document}

\maketitle

\newpage

\section{Introduction}\label{intro}

The notion of a {\it spectral curve\/} is one of the deep structures of an integrable system \cite{BBT}. At the classical level, the moduli of the curve are the action variables of the system, whilst the angle variables are associated to the Jacobian of the curve. Each classical solution determines a curve and the conserved charges of the integrable system can be read off the behaviour of a certain function---the quasi-momentum---defined on the curve at special points. There is an inverse algorithm for reconstructing the classical solution from a curve. If integrability survives quantization then it is natural to expect there to be a {\it Quantum Spectral Curve\/} (QSC). The curve provides a way to calculate the charges of states in the theory.

A fascinating integrable system is the world-sheet theory of the string in $\text{AdS}_5{\times} S^5$ and the ${\cal N}=4$ gauge-gravity duality.\footnote{The scope of this subject and the literature is vast. A key set of review articles are \cite{Beisert:2010jr}, another review is \cite{Arutyunov:2009ga}.} In this case, the form of the classical curve has been determined in \cite{Beisert:2005bm}. The curve is determined by the solution of some auxiliary integral equations that can be viewed as a complicated Riemann-Hilbert problem. Early proposals were then made for how the curve could be quantized leading to a set of rather novel Bethe Ansatz type equations, known as the Asymptotic Bethe Ansatz (ABA) \cite{Beisert:2005fw}. The ABA can be viewed as a direct quantization of the classical spectral curve and it
also had a direct connection to the dual gauge theory. The quantum spectral curve has been determined in \cite{Gromov:2013pga,Gromov:2014caa} (see also \cite{Gromov:2017blm}).

The $\text{AdS}_5{\times} S^5$ string theory admits various kinds of deformation that preserve integrability. In particular, there are the ``eta" \cite{Klimcik:2008eq,Delduc:2013fga,Delduc:2013qra}, ``beta" \cite{Lunin:2005jy,Kawaguchi:2014qwa,Osten:2016dvf} and ``lambda'' deformations \cite{Sfetsos:2013wia,Hollowood:2014rla,Hollowood:2014qma}. The QSC of the eta deformed string has been determined in \cite{Klabbers:2018tcd,Klabbers:2017vtw}. Here, we focus on the lambda deformation that for the bosonic sector can be viewed as the non-abelian T-duality of the theory with respect to the full $\text{PSU}(2,2|4)$ with a particular kind of compactification of the non-compact geometry that result from the na\"\i ve duality. This particular deformation has the necessary  kappa symmetries and is one-loop finite \cite{Appadu:2015nfa}.
The notion of a compactified non-abelian T-dual is crucial here in order that the na\"\i ve non-abelian T-dual is a consistent string background. It involves lifting the algebra-valued Lagrange multiplier field into a group-valued field. The lambda string model, has an associated level $k$ and in some sense, the na\"\i ve non-abelian T-dual is recovered in the limit $k\to\infty$.
The resulting $\lambda$ string is formulated as a gauged WZW model for the supergroup $F=\text{PSU}(2,2|4)$ where the full $F$ (vector) symmetry is gauged but with a deformation that breaks the gauge symmetry to the bosonic subgroup $G=\Sp(2,2){\times}\Sp(4)(\simeq\SO(1,4){\times}\SO(5))$. The fact that the lambda deformation gives rise to a consistent background for the string has been investigated in \cite{Sfetsos:2014cea,Demulder:2015lva,Borsato:2016zcf,Chervonyi:2016ajp,Borsato:2016ose}, with explicit results for AdS$_n\times S^n$ with $n=2$~\cite{Borsato:2016zcf} and 3~\cite{Chervonyi:2016ajp} and on general grounds for $n=5$ in \cite{Borsato:2016ose}. 

It is worth emphasizing that both the eta and lambda deformations are associated to a quantum group deformation of the symmetry algebra. For the former the deformation parameter $q$ is real while for the latter $q$ is a root of unity. The focus in the present work is on the lambda deformation and so we will only consider the case with $q$ a complex phase. There is nothing to prevent a similar analysis for the case with $q$ real but the details will be different and we do not present them here.

The quantum curve of the undeformed string model has been constructed by following a lengthy path that starts with the S-matrix of the excitations on the world-sheet and then proceeds via the TBA to the T-system and then beyond to the quantum spectral curve itself \cite{Gromov:2013pga}. For the lambda string model, the S-matrix for the 
world-sheet excitations has been determined in \cite{Hoare:2011wr}. This S-matrix is built from the $q$ deformed $R$-matrix of Beisert and Koroteev \cite{Beisert:2008tw} and has been tested in the semi-classical limit in \cite{Appadu:2017xku}. It should be possible to follow the same path to the quantum spectral curve for the lambda model. However, the form of the S-matrix and the resulting TBA, involving the quantum group at a root-of-unity, is much more opaque and it is by no means clear that this approach will be tractable. Another possible approach, and the one we adopt here and in a subsequent work \cite{qsc}, is to conjecture a form for the quantum curve based on the underlying symmetries and then show that it has the correct classical limit, i.e.~the classical spectral curve. Such an approach is not unreasonable because even the original approach via the TBA is not a first principles calculation (the S-matrix is a conjecture).

In the subsequent paper \cite{qsc}, the classical curve will then be seen to arise from the appropriate limit of the QSC, providing a way to test what amounts to an ansatz for the QSC. This limit is subtle but involves an intermediate stage where the QSC becomes a set of Bethe Ansatz equations for a spin chain. The fundamental difference is that the $\text{AdS}_5{\times} S^5$ case leads to an XXX type spin chain while the lambda string leads to an XXZ type chain. 

The organization of the paper is as follows. In section \ref{s10} we provide some background on Lie superalgebras and establish notation. In section \ref{s2} we provide a short review of the lambda string model paying particular attention to its integrability. Section \ref{s3} discusses the monodromy of the model from which the classical curve is constructed. We also argue how the charges extracted from the monodromy include the energy and momentum of the gauge-fixed worldsheet theory. In section \ref{s4}, we construct the classical spectral curve, writing it in terms of roots and weights of the superalgebra. Section \ref{s6} is devoted to showing that the classical curve can be identified with the semi-classical limit of a set of Bethe Ansatz equations. The details of this will be presented in the follow-up paper~\cite{qsc}. The appendix contains a discussion of the analogue of the light-cone gauge-fixed lambda superstring and in particular how the worldsheet Hamiltonian emerges.

\section{Some preliminaries}\label{s10}

In this section, we describe some aspects of the theory of Lie supergroups and algebras that we will need.

\subsection{Superalgebra Conventions}\label{a1}

The world-sheet theories of the conventional $AdS_5{\times} S^5$  superstring  and lambda deformation involve the semi-symmetric space $F/G=\text{PSU}(2,2|4)/\text{Sp}(2,2)\times\text{Sp}(4)$, whose bosonic part is precisely AdS$_5{\times} S^5$ realised as the product of symmetric spaces $\SU(2,2)/\Sp(2,2)\times\SU(4)/\Sp(4)$.
Our conventions for this superalgebra are  
taken from \cite{Arutyunov:2009ga}. First of all, the superalgebra ${\mathfrak sl}(4|4)$ is defined by the $8\times8$ matrices
\EQ{
M=\MAT{m&\theta\\ \eta&n}\ ,
}
where $m$ and $n$ are Grassmann even and $\theta$ and $\eta$ are Grassmann odd. 
These matrices are required to have vanishing supertrace\footnote{Notice that our convention is the opposite of \cite{Grigoriev:2007bu,Arutyunov:2009ga}, so that the supertrace is positive on the $S^5$ factor and negative on the $AdS_5$ factor.}
\EQ{
\text{str}\,M=-\Tr\,m+\Tr\,n=0\ .
\label{STr}
}
The non-compact real form $\msu(2,2|4)$ is picked out by imposing the reality condition
\EQ{
M=-HM^\dagger H
\label{real}
}
where, in $2\times 2$ block form,
\EQ{
H={\small\left(\begin{array}{cc|cc} -1_2 &&&\\ &1_2 &&\\ \hline &&1_2&
\\ &&&\ 1_2\ \end{array}\right)}\ .
\label{hh8}
}
Here, $\dagger$ is the usual hermitian conjugation, $M^\dagger=(M^*)^t$, but with the definition that complex conjugation is anti-linear on products of Grassmann odd elements
\EQ{
(\theta_1\theta_2)^*=\theta_2^*\theta_1^*\ ,
}
which guarantees that $(M_1M_2)^\dagger=M_2^\dagger M_1^\dagger$.
The superalgebra ${\mathfrak{psu}}(2,2|4)$ is then the quotient of $\mathfrak{su}(2,2|4)$ by the unit element $i{\mathbb I}_{8}$, which is the centre of the algebra. 

The algebra admits a $\mathbb Z_4$ automorphism $\sigma_-^4=1$ defined as
\EQ{
M\longrightarrow\sigma_-(M)=-{\cal K}M^{st}{\cal K}^{-1}\ ,
\label{dyy}
}
where $st$ denotes the ``super-transpose'' defined as
\EQ{
M^{st}=\MAT{m^t&-\eta^t\\ \theta^t&n^t}\ .
}
and 
\EQ{
{\cal K}={\small\left(\begin{array}{cc|cc} J_2 &  &  & \\
  & J_2 &  & \\ \hline
 &  & J_2 & \\
 &  & & J_2\end{array}\right)}\ ,\qquad J_2=\MAT{0&-1\\ 1&0}\ .
}

Under $\sigma_-$, the superalgebra ${\mathfrak{psu}}(2,2|4)$ has the decomposition
\EQ{
{\mathfrak f}= {\mathfrak f}^{(0)}\oplus{\mathfrak f}^{(1)}\oplus{\mathfrak f}^{(2)}\oplus{\mathfrak f}^{(3)}\ ,\qquad \sigma_-({\mathfrak f}^{(j)})=i^j\,{\mathfrak f}^{(j)}\ , \qquad [{\mathfrak f}^{(j)},{\mathfrak f}^{(k)}]\subset {\mathfrak f}^{(j+k\; \text{mod}\; 4)}\,.
\label{CanonicalDecN}
}
In particular, the even graded parts are Grassmann even while the odd graded parts are Grassmann odd. The zero graded part ${\mathfrak f}^{(0)}\equiv{\mathfrak g}$ is the (bosonic) Lie algebra
of $G=\Sp(2,2)\times\Sp(4)$. Correspondingly, ${\mathfrak f}^{(0)}\oplus{\mathfrak f}^{(2)}$ is the Lie algebra of $\SU(2,2)\times\SU(4)$, which is the bosonic subgroup of $F$.
Moreover, $\STr(ab)=0$ for any $a\in\mathfrak{f}^{(i)}$, $b\in\mathfrak{f}^{(j)}$ with $i+j\not=0\ \text{mod}\ 4$.

In the defining representation, the generators are matrices $E_{AB}$ with a 1 in position $(A,B)$ and zeros elsewhere. We can label the generators in the same way in an arbitrary representation. Each of the labels $A\in\{1,2,\ldots,8\}$ can be assigned a $p$-parity $p_A\in\{0,1\}$ and a $c$-parity $c_A\in\{0,1\}$. The $p$-parity determines the graded Lie bracket of 2 generators:
\EQ{
[E_{AB},E_{CD}]=\delta_{BC}E_{AD}-(-1)^{(p_A+p_B)(p_C+p_D)}\delta_{AD}E_{CB}\ .
}
So generators with $p_A+p_B$ even/odd are even/odd generators of the Lie superalgebra. In our case, we have $p_A=(1,1,1,1,0,0,0,0)$.
The $c$-parity determines the reality condition on the generators
by specifying an anti-linear conjugation of the form
\EQ{
E_{AB}^\ddagger=(-1)^{c_A+c_B}E_{BA}\ .
}
The real form is then defined by taking anti-Hermitian combinations with respect to this definition of the conjugate. 
Writing~\eqref{real} as $M=-M^\ddagger$, it follows that in our case $c_A=(1,1,0,0,0,0,0,0)$.

\subsection{Roots and weights}

The Cartan subspace of the superalgebra in the defining representation consist of the diagonal matrices $E_{AA}$. These are all even generators. We can introduce a root/weight vector space spanned by a set of vectors $\{\Be_A\}=\{\Be_i,\Bdelta_i\}$ with
\EQ{
\Be_i\cdot\Be_j=\delta_{ij}\ ,\qquad\Be_i\cdot\Bdelta_j=0\ ,\qquad\Bdelta_i\cdot\Bdelta_j=-\delta_{ij}\ .
}
In a Cartan-Weyl basis, we can think of the Cartan generators as a vector
\EQ{
\BH=\sum_i\big(\Bdelta_iE_{ii}+\Be_iE_{4+i,4+i}\big)\ ,
}
so that $[\BH,E_\Balpha]=\Balpha E_\Balpha$. This identifies, for $i\neq j$, the even generators and roots
\EQ{
E_{ij}=E_{\Bdelta_i-\Bdelta_j}\ ,\qquad E_{i+4,j+4}=E_{\Be_i-\Be_j}\ ,
}
and then, the odd generators and roots,
\EQ{
E_{i,j+4}=E_{\Bdelta_i-\Be_j}\ ,\qquad E_{i+4,j}=E_{\Be_i-\Bdelta_j}\ .
}

A weight vector is then expressed as 
\EQ{
\Bomega=\sum_{i=1}^4\big(\nu_i\Bdelta_i+\lambda_i\Be_i\big)\ ,
}
so that
\EQ{
\STr\big((\Bomega\cdot\BH)(\Bomega'\cdot\BH)\big)=\Bomega\cdot \Bomega'\ .
}

We can extend the action of the $\mathbb Z_4$ automorphism $\sigma_-$ onto weight vectors,
\EQ{
\sigma_-(\Be_1)&=-\Be_2\ ,\qquad \sigma_-(\Be_2)=-\Be_1\ ,\\
\sigma_-(\Be_3)&=-\Be_4\ ,\qquad \sigma_-(\Be_4)=-\Be_3\ ,\\
\sigma_-(\Bdelta_1)&=-\Bdelta_2\ ,\qquad \sigma_-(\Bdelta_2)=-\Bdelta_1\ ,\\
\sigma_-(\Bdelta_3)&=-\Bdelta_4\ ,\qquad \sigma_-(\Bdelta_4)=-\Bdelta_3\ .
\label{zz13}
}
Note that on this vector space $\sigma_-$ has order 2. 

The Cartan subalgebra of $\mathfrak{psu}(2,2|4)$ is six dimensional because of two properties. Firstly the unimodular property, the $\mathfrak s$ in $\mathfrak{psu}$, implies that it only includes 7 independent elements which, in the defining representation are a basis of supertraceless diagonal matrices. For example, in a general representation we can take elements
\EQ{
i(\Bdelta_j-\Bdelta_{j+1})\cdot\BH\ ,\qquad i(\Be_j-\Be_{j+1})\cdot\BH\ ,\qquad i(\Bdelta_4-\Be_1)\cdot\BH\ ,
}
for $j=1,2,3$. There is a central term 
\EQ{
\mathfrak C=i\sum_{A=1}^8E_{AA}=i\sum_{i=1}^4 \big(-\Bdelta_i +\Be_i\big)\cdot \BH\ ,
}
which commutes with the rest of the algebra. The projectivity property, the $\mathfrak p$ in $\mathfrak{psu}$, implies the equivalence relation $a+\mathfrak C\sim a$, for any element of the algebra $a$. It follows that in a particular representation, the central term must vanish which implies that weight vectors satisfy the constraint
\EQ{
\sum_{i=1}^4 \big(-\Bdelta_i +\Be_i\big)\cdot \Bomega=\sum_{i=1}^4\big(\nu_i+\lambda_i\big)=0\ .
\label{ccc}
}

Of the remaining six elements in the Cartan subalgebra, four are in $\mf^{(0)}$ and two are in $\mf^{(2)}$. The two elements in $\mf^{(2)}$ are associated to vectors that are odd under $\sigma_-$,
\EQ{
\Bxi_1=\Bdelta_1+\Bdelta_2-\Bdelta_3-\Bdelta_4\ ,\qquad
\Bxi_2=\Be_1+\Be_2-\Be_3-\Be_4\ ,\qquad \ .
\label{lip}
}
The associated generators in the defining representation are
\EQ{
\Lambda_1\equiv -\frac i2\Bxi_1\cdot\BH=\frac i2{\small\left(\begin{array}{cc|cc} 1_2 &  &  & \\
  & ~-1_2 &  & \\ \hline
 &  & ~0_2 & \\
 &  & & ~0_2\end{array}\right)}\ ,\qquad \Lambda_2\equiv \frac i2\Bxi_2\cdot\BH=\frac i2{\small\left(\begin{array}{cc|cc} ~0_2 &  &  & \\
  & ~0_2 &  & \\ \hline
 &  & ~1_2 & \\
 &  & & -1_2\end{array}\right)}\ .
 \label{zzl}
 }

In a conventional Lie algebra, all choices of simple roots are identical up to the action of the Weyl group. In a Lie superalgebra, there are no Weyl reflections associated to the odd roots and so there are inequivalent choices of simple roots. For example, one particular choice that we will use extensively is associated to the ordering
\EQ{
\Bdelta_1\ ,\quad \Be_1\ ,\quad\Be_2\ ,\quad \Bdelta_2\ ,\quad \Bdelta_3\ ,\quad \Be_3\ ,\quad\Be_4\ ,\quad \Bdelta_4\ .
\label{hd6}
}
The simple roots are then the differences of adjacent vectors in the ordering and can be associated to the Kac-Dynkin-Vogan (KDV) diagram
\begin{equation}
\begin{tikzpicture}[scale=0.7]
\node at (-1,1) {$\Bdelta_1$};
\node at (1,1) {$\Be_1$};
\node at (3,1) {$\Be_2$};
\node at (5,1) {$\Bdelta_2$};
\node (a1) at (7,1) {$\Bdelta_3$};
\node (a2) at (9,1) {$\Be_3$};
\node at (11,1) {$\Be_4$};
\node at (13,1) {$\Bdelta_4$};
\node at (8,-1.5) {$\Balpha_5=\Bdelta_3-\Be_3$};
\draw[thin,->] (7,-1.2) -- (7.6,-0.5);
\draw[thin,->] (a1) to[out=-90,in=130] (7.8,-1.1);
\draw[thin,->] (a2) to[out=-90,in=80] (9.3,-1.1);
\draw (0.5,0) -- (1.5,0);
\draw (2.5,0) -- (3.5,0);
\draw (4.5,0) -- (5.5,0);
\draw (6.5,0) -- (7.5,0);
\draw (8.5,0) -- (9.5,0);
\draw (10.5,0) -- (11.5,0);
\draw (2,0) circle (0.5cm);
\draw (6,0) circle (0.5cm);
\draw (6,0) circle (0.4cm);
\draw (10,0) circle (0.5cm);
\begin{scope}[xshift=0cm]
\draw (0,0) circle (0.5cm);
\draw (0,0) circle (0.4cm);
\draw (0.285,0.285) -- (-0.285,-0.285);
\draw (0.285,-0.285) -- (-0.285,0.285);
\end{scope}
\begin{scope}[xshift=4cm]
\draw (0,0) circle (0.5cm);
\draw (0,0) circle (0.4cm);
\draw (0.285,0.285) -- (-0.285,-0.285);
\draw (0.285,-0.285) -- (-0.285,0.285);
\end{scope}
\begin{scope}[xshift=8cm]
\draw (0,0) circle (0.5cm);
\draw (0.353,0.353) -- (-0.353,-0.353);
\draw (0.353,-0.353) -- (-0.353,0.353);
\end{scope}
\begin{scope}[xshift=12cm]
\draw (0,0) circle (0.5cm);
\draw (0.353,0.353) -- (-0.353,-0.353);
\draw (0.353,-0.353) -- (-0.353,0.353);
\end{scope}
\end{tikzpicture}
\label{bn1}
\end{equation}
A KDV diagram encodes the properties of the simple roots. In particular, the fermionic simple roots are indicated with a cross; these are null $\Balpha^2=0$. Bosonic roots have either $\Balpha^2=\pm2$. The double circles encode the non-compactness of the Lie supergroup in the sense that a simple root with a $c$-parity odd generator is indicated by a double circle on the KDV diagram. In the notation of weights and with our choice of simple roots, the $p$ and $c$ parities are
\EQ{
&p_{\Bdelta_i}=1\ ,\qquad p_{\Be_i}=0\ ,\\
&c_{\Bdelta_1}=c_{\Bdelta_2}=1\ ,\qquad c_{\Bdelta_3}=c_{\Bdelta_4}=c_{\Be_i}=0\ ,
}
which then defines the parities of roots, e.g.~$p_{\Bdelta_1-\Be_1}=p_{\Bdelta_1}+p_{\Be_1}$, etc.

\subsection{The loop algebra}

Associated to the Lie superalgebra $\mathfrak{psu}(2,2|4)$ is a twisted affine (loop) algebra
\EQ{
\hat\mf=\bigoplus_{n\in\mathbb Z}\Big(\bigoplus_{j=0}^3\mf^{(j)}\otimes z^{4n+j}\Big)\ ,
\label{twa}
} 
where $z$ is the spectral parameter. Taking into account \eqref{real} and \eqref{dyy}, we see that a loop superalgebra element $a(z)\in \hat\mf$ satisfies
\EQ{
a(iz)=\sigma_-\big(a(z)\big)=- {\cal K}a^{st}(z) {\cal K}^{-1}\,,
\qquad
a(z^\ast)=-Ha^\dagger(z) H\,.
}

\section{The lambda string}
\label{s2}

In this section, we present a mini review of the worldsheet sigma model for the string on AdS$_5\times S^5$ and its associated 
 lambda model deformation in conformal gauge, where the world-sheet metric is of the form $h_{\mu\nu}= e^\phi\, \text{diag}(1,-1)$. 
The undeformed model involves a coset $F/G$, $F=\text{PSU}(2,2|4)$ and $G=\text{Sp}(2,2)\times\text{Sp}(4)$, with a particular WZ term that is exact and so can be written in local form \cite{Metsaev:1998it}. The field $f(\tau,\sigma)$ is valued in the Lie supergroup $F$  and the current $J_\mu=f^{-1}\partial_\mu f$ takes values in its Lie superalgebra $\mf$. The action takes the form\footnote{In our notation $\sigma^\pm=\tau\pm\sigma$ and $\partial_\pm=(\partial_0\pm\partial_1)/2$ and so for vectors $A^\pm=A^0\pm A^1$ and $A_\pm=(A_0\pm A_1)/2$. The 
Weyl invariant combination of the world-sheet metric $\gamma^{\mu\nu}= \sqrt{-\text{det}(h)}\, h^{\mu\nu}$
has components $\gamma^{00}=-\gamma^{11}=1$, and $\varepsilon^{01}=-\varepsilon^{10}=1$, so that $\gamma^{+-}=\gamma^{-+}=\varepsilon^{-+}=-\varepsilon^{+-}=2$. We shall also use $d^2\sigma\equiv d\tau\, d\sigma$.}
\EQ{
S_\sigma[f]=-\frac{\kappa^2}\pi\int d^2\sigma\,\STr\Big[J_+^{(2)}J^{(2)}_-
+\frac12J^{(1)}_+J^{(3)}_- -\frac12J^{(1)}_-J^{(3)}_+\Big]\ .
\label{pss}
}
Here, the superscripts correspond to the decomposition into the eigenspaces of the $\mathbb Z_4$ automorphism of the Lie superalgebra, $\mathfrak f=\oplus_{j=0}^3\mathfrak f^{(j)}$. We will define projectors $\mathbb P^{(j)}$ onto these eigenspaces. The invariant subspace $\mathfrak f^{(0)}=\mathfrak g$ is the Lie algebra of the bosonic subgroup $G$.
This action is invariant under global $F_L$ transformations
\EQ{
f\to Uf \,,\qquad U\in F \,,
}
which leave the current $J_\mu$ invariant. 
The equations-of-motion of the undeformed model, along with the Cartan-Maurer identity $\partial_+J_--\partial_-J_++[J_+,J_-]=0$, can be written in Lax form
\EQ{
[\partial_\mu+\LAX_\mu(z),\partial_\nu+\LAX_\nu(z)]=0\ ,
\label{leq2}
}
where\footnote{We use the notation of \cite{Appadu:2017xku} related to that of \cite{Hollowood:2014qma} by $z\to1/z$ and $\mathfrak f^{(1)}\leftrightarrow\mathfrak f^{(3)}$.}
\EQ{
\LAX_\pm(z)=J_\pm^{(0)}+z J_\pm^{(1)}+z^{\pm2} J_\pm^{(2)}+z^{-1} J_\pm^{(3)}\ ,
\label{rmm}
}
and $z$ is the arbitrary spectral parameter. Notice that the Lax connection $\LAX_\mu(z)$ takes values in the twisted loop algebra \eqref{twa}, with the spectral parameter $z$ being the affine parameter.

The lambda model is based on a WZW model for an
$F$-valued group field $\CF$ along with a gauge field $A_\mu$ valued in  $\mathfrak f$ \cite{Hollowood:2014qma}. Actually, only the zero graded component of $A_\mu$ is a genuine gauge field while the other components are Gaussian auxiliary fields. The action takes the form
\EQ{
S_\lambda[\CF ,A_\mu]=S_\text{gWZW}[\CF ,A_\mu]-\frac {k}{\pi}\int d^2\sigma\,\STr\,\big[
A_+\big(\Omega_+-1\big)A_-\big]
\label{dWZW}
}
where
\EQ{
S_\text{gWZW}[\CF ,A_\mu]&=-\frac k{2\pi}\int d^2\sigma\STr\Big[
\CF ^{-1}\partial_+\CF \,\CF ^{-1}\partial_-\CF +2A_+\partial_-\CF \CF ^{-1}\\ &~~~~~~~~~
-2A_-\CF ^{-1}\partial_+\CF -2\CF ^{-1}A_+\CF  A_-+2A_+A_-\Big]
\\ &~~~~~~~~~+\frac k{12\pi}\int d^3y\,\epsilon^{abc}\Tr\,\Big[\CF ^{-1}\partial_a\CF \,
\CF ^{-1}\partial_b\CF \,\CF ^{-1}\partial_c\CF \Big]\ ,
\label{gWZW}
}
is the gauged WZW action for $F$ gauged by the action $\CF\to U\CF U^{-1}$ for $U\in F$ \cite{Karabali:1988au,Gawedzki:1988hq,Karabali:1989dk},
and
\EQ{
\Omega_\pm=\PP^{(0)}+\lambda^{\pm1}\PP^{(1)}+\lambda^{-2}\PP^{(2)}+\lambda^{\mp1}\PP^{(3)}\ .
}
Notice that $\Omega_\pm-1$ vanishes on $\mf^{(0)}$ manifesting the fact that the gauge symmetry of $S_\lambda$ corresponds to the  bosonic subgroup $G\subset F$.

The equations of motion for $A_\mu$ give the constraints
\EQ{
\CF ^{-1}\partial_+\CF +\CF ^{-1}A_+\CF&=\Omega_-A_+\ ,\\
\partial_-\CF \CF ^{-1}-\CF A_-\CF ^{-1}&=-\Omega_+ A_-\ .
\label{cox}
}
They can be written in terms of the usual Kac-Moody currents of the gauge WZW model
\EQ{
\JJ_\pm=-\frac k{2\pi}\big(\CF ^{\mp1}\partial_\pm\CF^{\pm1} +\CF ^{\mp1}A_\pm\CF^{\pm1} -A_\mp\big)
\label{kmc}
}
as follows
\EQ{
\JJ_\pm+\frac k{2\pi}\big(\Omega_\mp A_\pm-A_\mp\big)=0\ .
}
The components of these constraints in $\mathfrak f^{(1)}\oplus \mathfrak f^{(2)}\oplus
\mathfrak f^{(3)}$ are second class whereas the constraints in $\mathfrak f^{(0)}$ are first class.
The equations of motion of the group field can be written as either as the following
\EQ{
&[\partial_++\CF ^{-1}\partial_+\CF +\CF ^{-1}A_+\CF,\partial_-+A_-]=0\ ,\\
&[\partial_++A_+,\partial_--\partial_-\CF\CF^{-1}+\CF A_-\CF^{-1}]=0\ ,
\label{sap}
}
which are trivially equivalent by conjugation with $\CF$. 

Since the Lagrangian action is quadratic in the fields $A_\mu$, imposing eqs.~\eqref{cox} for the components $A_\mu^{(j)}$, $j=1,2,3$ (i.e.~the non-zero graded pieces of $A_\mu$), is equivalent to integrating out those fields at the classical level. 
Then, the equations of motion of~$\CF$ in the lambda model can be written in the same Lax form as the undeformed model \eqref{leq2} and~\eqref{rmm} with
\EQ{
J_\pm=A^{(0)}_\pm+\lambda^{\mp1/2}A^{(1)}_\pm+\lambda^{-1}A^{(2)}_\pm+\lambda^{\pm1/2}A^{(3)}_\pm\ .
\label{LaxLM}
}
Integrating out the fields $A_\mu$ (after suitable gauge fixing) allows one to write them in terms of $\CF$ as follows
\EQ{
A_+=-\big(\text{Ad}_{\CF^{-1}}-\Omega_-\big)^{-1}\, \CF^{-1}\partial_+\CF\,,\qquad
A_-=\big(1-\text{Ad}_{\CF^{-1}} \Omega_+\big)^{-1}\, \CF^{-1}\partial_-\CF\,.
}
This gives rise to an effective sigma model for the field $\CF$ that can be understood as a deformation of the non-abelian T-dual of the sigma model with respect to its full $\text{PSU}(2,2|4)$ global symmetry group \cite{Hollowood:2014qma}. 
It is important to notice that, at the quantum level, the super-determinant that arises in integrating out $A_\mu$ produces a dilaton on the world-sheet.

It is worth pointing out the relation between the original undeformed model, a sigma model on the semi-symmetric space, and the lambda model. This involves the following steps which we describe in reverse starting from the lambda model \cite{Hollowood:2014qma}. The idea is to take the combined limit 
\EQ{
k\to\infty\ ,\qquad\lambda\to1\ ,\quad\text{with}\quad
\kappa^2\equiv2(1-\lambda)k\quad  \text{fixed}\ ,
\label{liml}
}
and expanding the group field around the identity ${\cal F}=1+\kappa^2\nu/k+\cdots$, in which case the gauged WZW action reduces to
\EQ{
S_\text{gWZW}[{\cal F},A_\mu]\rightarrow -\frac{\kappa^2}\pi\int d^2\sigma\,\STr(\nu F_{+-})\ ,
}
where $F_{\mu\nu}$ is the field strength of the connection $A_\mu$. The second term in 
\eqref{dWZW} reduces to the original action \eqref{pss} with $J_\mu$ replaced by $A_\mu$. The field $\nu$ acts as a Lagrange multiplier that imposes the flatness of $A_\mu$ which implies that there exits a group field $f$ such that $A_\mu=f^{-1}\partial_\mu f$. We have, therefore, recovered the original undeformed model in this particular limit.

\subsection{Virasoro constraints}

In both the undeformed and lambda models, we need to impose the Virasoro constraints, 
\EQ{
\STr(J_+^{(2)}J_+^{(2)})=\STr(J_-^{(2)}J_-^{(2)})=0\ ,
\label{epk}
}
that are left over after fixing conformal gauge. These conditions are solved by requiring
\EQ{
J^{(2)}_\pm=\mu_\pm V_\pm\Lambda V_\pm^{-1}\ ,
\label{jla}
}
where $\mu_\pm$ are arbitrary functions and $V_\pm\in G$, the group associated to the zero graded subalgebra, so that having $\Lambda\in\mf^{(2)}$ implies that the current components are valued in $\mf^{(2)}$. In the above, we have defined 
\EQ{
\Lambda= \Lambda_1+\Lambda_2\ ,
\label{zz2}
}
where we defined $\Lambda_1$ and $\Lambda_2$ in \eqref{zzl}.
Note that we could equally as well have chosen the combination $\Lambda_1-\Lambda_2$ but this is conjugate in $G$ to our choice above.

There is a second type of solution of the form
\EQ{
\Lambda={\small\left(\begin{array}{cc|cc} i1_2 & 1_2 &  & \\
1_2  & -i1_2 &  & \\ \hline
 &  & ~0~& \\
 &  & & ~0~\end{array}\right)}\in \mathfrak{f}^{(2)}\ ,
\label{ldd3}
}
which takes values in the subalgebra $\msu(2,2)^{(2)}$, and gives rise to string configurations that stay entirely in $\text{AdS}_5$~\cite{Tseytlin:2010jv}. It is not clear the r\^ole that this sector plays in the ordinary string, let alone the lambda deformation discussed here, and so we will not discuss this second type of solutions any further. 

\subsection{Gauge symmetries}

The Lagrangian action of the undeformed and lambda models exhibit three different gauge symmetries each. First of all, the two models are invariant under conformal world-sheet reparameterizations $\sigma^\pm\to \widetilde{\sigma}^\pm(\sigma^\pm)$, which are not fixed by the conformal gauge condition.
In addition, the undeformed model action~\eqref{pss} is invariant under (infinitesimal) $G_R$ gauge transformations
\EQ{
\delta f = -fu^{(0)}\,,\qquad u^{(0)}\in \mathfrak{f}^{(0)}\,,
\label{GRgauge}
}
and under kappa symmetry transformations
\EQ{
\delta f= -f \epsilon\,,
\label{kappaSM}
}
where $\epsilon$ is of the form
\EQ{
\epsilon=\epsilon^{(1)}+\epsilon^{(3)}=\big[J_+^{(2)}, \kappa^{(1)}\big]_+ +\big[J_-^{(2)}, \kappa^{(3)}\big]_+\in \mathfrak{f}^{(1)}\oplus \mathfrak{f}^{(3)}\,.
\label{fparam}
}
Here, $\kappa^{(1)}\in \mathfrak{f}^{(1)}$ and $\kappa^{(3)}\in \mathfrak{f}^{(3)}$ are infinitesimal (fermionic) local parameters, and $[a,b]_+=ab+ba$ denotes the anti-commutator.

Similarly, the lambda model action is invariant under $G_V$ gauge transformations
\EQ{
\delta \CF= [u^{(0)}, \CF]\,,\qquad \delta A_\pm =-[\partial_\pm +A_\pm,u^{(0)}] \,,\qquad u^{(0)}\in \mathfrak{f}^{(0)}\,.
\label{GVgauge}
}
Moreover, it is also invariant under the kappa symmetry transformations~\cite{Hollowood:2014qma}
\EQ{
\delta\CF= \alpha \CF -\CF \beta\,,\qquad \delta A_+= -[\partial_+ + A_+, \alpha]
\,,\qquad \delta A_-= -[\partial_- + A_-, \beta]\,,
\label{kappaLM}
}
where
\EQ{
\alpha=\lambda^{1/2} \epsilon^{(1)} + \lambda^{-1/2} \epsilon^{(3)} \,,\qquad
\beta=\Omega_-\alpha=\lambda^{-1/2} \epsilon^{(1)} + \lambda^{1/2} \epsilon^{(3)}
}
and $\epsilon^{(1)}$ and $\epsilon^{(3)}$ are of the form~\eqref{fparam}.\footnote{Note that this transformation is a symmetry of the Lagrangian action in the conformal gauge provided that one imposes the Virasoro constraints. In general, the kappa symmetry transformation involves a specific variation of the world-sheet metric.}

\subsection{The wave function}

The fact that the sigma and lambda models share the same Lax 
connection provides a map between the solutions to the equations of motion of the two models, including the Virasoro constraints. This map becomes explicit by introducing the ``wave function'' to which we now turn.

The zero-curvature condition~\eqref{leq2} is the compatibility condition of the associated linear problem
\EQ{
\big[\partial_\mu+{{\LAX}}_\mu(z)\big]\Psi(z)=0\ ,
\label{lprob}
}
whose solution is the wave function $\Psi(\tau,\sigma;z)\equiv \Psi(z)$. It takes values in the loop group $\widehat F$ associated to the twisted loop algebra~\eqref{twa},  so it satisfies
\EQ{
{\cal K}\, \Psi^{st}(z)\, {\cal K}^{-1}=\Psi^{-1}(iz)\ ,\qquad H\Psi^\dagger(z) H=\Psi^{-1}(z^*)\ .
}
Notice that the linear problem~\eqref{lprob} determines the 
wave function up to multiplication on the right by a constant element of $\widehat{F}$\footnote{It is customary to fix this freedom by imposing an initial condition like $\Psi(0,0;z)=1$. However, we prefer to leave this choice free in the following.}
\EQ{
\Psi(\tau,\sigma;z)\to\Psi(\tau,\sigma;z)\,g(z)\,, \qquad g(z)\in \widehat{F}\ .
\label{redwave}
}

The wave function is defined on-shell and contains all the information about the space of solutions to the equations of motion. In terms of $\Psi(z)$, the Lax connection can be recovered from
\EQ{
\LAX_\mu(z)=-\partial_\mu \Psi(z)\Psi^{-1}(z)\,.
}
Moreover, one can write the group valued fields of both the undeformed and the lambda models in terms of $\Psi(z)$. Namely, using~\eqref{rmm} and the definition of $J_\mu$ in terms of the field $f$, it follows that $\LAX_\mu(1)=f^{-1}\partial_\mu f$ and so
\EQ{
\big(\partial_\mu + {{\LAX}}_\mu(1)\big)\, f^{-1}=0\ .
}
Thus, 
\EQ{
f=\Psi^{-1}(1)\,.
\label{groupel}
}
There is the freedom to perform global right transformations \
\EQ{
\Psi(z)\to \Psi(z)V\ ,
\label{freet}
}
for $V\in F$, which exhibits the invariance of the equations of motion, and of the Lagrangian action in this case, under global $F_L$ transformations $f\to V^{-1}f$.

Similarly, using the identities
\EQ{
{{\LAX}}_{\pm}(\lambda^{\pm1/2})=A_\pm\ ,
\quad
{{\LAX}}_\pm(\lambda^{\mp1/2})=\Omega_\mp A_\pm\ ,
}
the equations of motion for $A_\mu$~\eqref{cox} 
become simply
\EQ{
\partial_\mu\CF =-{{\LAX}}_\mu(\lambda^{1/2})\CF +\CF {{\LAX}}_\mu(\lambda^{-1/2})\,,
}
which implies 
\EQ{
\CF=\Psi(\lambda^{1/2})\, \Psi^{-1}(\lambda^{-1/2})\,.
\label{groupelb}
}
The freedom to perform the global transformations \eqref{freet} now corresponds to a global symmetry of the equations of motion but not a manifest symmetry of the Lagrangian action; however, it was argued in~\cite{Appadu:2017xku} that the action of the lambda model in the conformal gauge exhibits off-shell global symmetries that are precisely of this form. 

The wave function allows one to describe in a unified way the action of the gauge symmetries of the sigma and lambda model on the space of solutions to their equations of motion. Namely, the $G_R$ and $G_V$ gauge transformations~\eqref{GRgauge} and~\eqref{GVgauge} correspond to
\EQ{
\delta_G \Psi(z)= u^{(0)} \Psi(z)\,.
\label{GPsi}
}
Moreover, in terms of the wave function, the kappa symmetry transformations~\eqref{kappaSM} and~\eqref{kappaLM} of the sigma and lambda models become simply
\EQ{
\delta_\kappa \Psi(z)= \big(z\epsilon^{(1)} + z^{-1}\epsilon^{(3)} \big) \Psi(z)\,.
\label{kappaPsi}
}

\section{Monodromy and symplectic form}\label{s3}

In this section, we define the monodromy of the integrable system underlying the sigma and lambda models. This object is key for going on to define the CSC.

We assume that $-\pi\leq \sigma\leq \pi$ and define the monodromy of the Lax connection as
\EQ{
T(\tau;z)&=\Psi(\tau,\pi;z) \Psi^{-1}(\tau,-\pi;z)\\[5pt]
&=\overset{\longleftarrow}{\text{Pexp}}\left[-\int_{-\pi}^{\pi}\, d\sigma\, \LAX_1(\tau,\sigma;z)\right]\equiv T(z)\,.
\label{monodromy}
}
This satisfies
\EQ{
\partial_0T(z)= -\LAX_0(\tau,\pi;z)\,T(z) +T(z)\,\LAX_0(\tau,-\pi;z)\ .
}
Hence, if we impose closed string boundary conditions on the Lax connection,\footnote{In this work, we will restrict ourselves to the case of closed strings. The discussion of open strings requires a modified definition of the monodromy.}
\EQ{
\LAX_0(\tau,\pi;z)=\LAX_0(\tau,-\pi;z)\ ,
\label{pLx}
}
the spectrum of the monodromy $T(\tau;z)$ is conserved in time.  Since $z$ is a free parameter this means that there are an infinite number of conserved quantities as required in an integrable field theory. 
In particular, if either the undeformed model field~$f$ or the lambda model field~$\CF$ are periodic functions of $\sigma$
\EQ{
f(\tau,\sigma)=f(\tau,\sigma+2\pi)\quad \text{or} \quad \CF(\tau,\sigma)=\CF(\tau,\sigma+2\pi)\ ,
}
then \eqref{pLx} is satisfied. 

It is worth noticing that the monodromy is invariant under the transformation~\eqref{redwave}. Moreover, under the gauge transformations~\eqref{GPsi} and~\eqref{kappaPsi}, which include the kappa symmetry transformations, it changes as follows
\EQ{
&\delta_G T(\tau;z)= u^{(0)}(\tau,\pi) T(\tau;z) -T(\tau;z) u^{(0)}(\tau,-\pi)\,,\\[5pt]
&\delta_\kappa T(\tau;z)=\big[z\epsilon^{(1)}  + z^{-1}\epsilon^{(3)} \big]_{\sigma=\pi} T(\tau;z) -T(\tau;z) \big[z\epsilon^{(1)}  + z^{-1}\epsilon^{(3)} \big]_{\sigma=-\pi} \,.
}
This shows that the spectrum of $T(\tau;z)$ is gauge invariant, provided that the local parameters $u^{(0)}$, $\kappa^{(1)}$, and $\kappa^{(3)}$ are also periodic.

Once we have the monodromy, we can define the Classical Spectral Curve (CSC), the algebraic curve defined by the characteristic equation of the monodromy:
\EQ{
F(p,z)\equiv\det\big[e^{ip}{\bf 1}-T(z)\big]=0\ .
\label{d33}
}
The eigenvalues $e^{ip_A(z)}$ of $T(z)$, $A=1,2,\ldots,8$, thought of as a function of $z$, define the quasi-momenta $p_A(z)$ and correspond to the branches of the function $p(z)$ that touch at various branch points. In the following, we will label the eigenvalues of the monodromy and the quasi-momentum as
\EQ{
T(z)\longrightarrow\text{diag}\Big(\underbrace{e^{i\hat p_1(z)},e^{i\hat p_2(z)},e^{i\hat p_3(z)},e^{i\hat p_4(z)}}_{\SU(2,2)}\Big|\underbrace {e^{i\tilde p_1(z)},e^{i\tilde p_2(z)},e^{i\tilde p_3(z)},e^{i\tilde p_4(z)}}_{\SU(4)}\Big)\ .
\label{ttr}
}
The quasi-momenta generate all the conserved quantities in the theory. Some of them are local in the fields of the theory, as we shall see, whereas the remainder are non-local. We will often think of the quasi-momenta as a Cartan vector
\EQ{
\Bp(z)=\sum_{i=1}^4\big(-\hat p_i(z)\Bdelta_i+\tilde p_i(z)\Be_i\big)\ ,
\label{goo}
}
defined so that 
\EQ{
p(z)\equiv\Bp(z)\cdot\BH=\sum_{i=1}^4\big(\hat p_i(z)E_{ii}+\tilde p_i(z)E_{4+i,4+i}\big)
}
to be consistent with \eqref{ttr}. Note the presence of the minus sign in \eqref{goo} which is arises because of the indefinite signature of the Cartan vector space.

The periodic boundary conditions~\eqref{pLx} ensure that $\Psi(\tau,\pi;z)= \Psi(\tau,-\pi;z){\cal W}(z)$, where ${\cal W}(z)$ is a constant element of the loop group. Since
\EQ{
T(\tau;z)=\Psi(\tau,-\pi;z)\,{\cal W}(z)\, \Psi^{-1}(\tau,-\pi;z)\,,
}
${\cal W}(z)$ and $T(\tau;z)$ have the same eigenvalues and, thus, one could equivalently define the CSC in terms of ${\cal W}(z)$.

The undeformed and lambda models share the same Lax connection and, therefore, one suspects that both models share the same CSC. However, this turns out not to be the case due to the way that closed string boundary conditions are imposed on the curves.
Note that the boundary conditions~\eqref{pLx} do not imply that
the Lagrangian fields $f$ and $\CF$ are periodic. For the undeformed model, using~\eqref{groupel}, one can easily show that
\EQ{
T(1)=f^{-1}(\tau,\pi) f(\tau,-\pi)\,.
}
Therefore, imposing closed string boundary conditions $f(\tau,\pi)=f(\tau,-\pi)$ is equivalent to $T(1)=1$. This is related to the well-known level-matching condition, which relates periodic boundary conditions to the vanishing of the total world-sheet momentum~\cite{Arutyunov:2009ga}.

Similarly, using~\eqref{groupelb}, the lambda model field satisfies
\EQ{
T(\lambda^{-1/2})=\CF^{-1}(\tau,\pi)\, T(\lambda^{1/2})\, \CF(\tau,-\pi)\,.
}
Hence, in this case, imposing closed string boundary conditions on $\CF$,
\EQ{
\CF(\tau,\pi)=\CF(\tau,-\pi)\,,
}
implies that the monodromy at the two special points $T(\lambda^{1/2})$ and $T(\lambda^{-1/2})$ have the same spectrum of eigenvalues. This will be the lambda string version of the level matching condition:
\EQ{
\{p_A(\lambda^{1/2})\}=\mathbb P\{p_A(\lambda^{-1/2})\}\ .
\label{lmc}
}
Here, $\mathbb P$ is a possible permutation that arises from the freedom afforded by the Weyl group acting on the Cartan subalgebra of the loop group. The allowed permutations, i.e.~ones that preserve the grading, are generated by the five elements $\Be_1\leftrightarrow\Be_2$, $\Be_3\leftrightarrow\Be_4$, $(\Be_1\leftrightarrow\Be_3,\Be_2\leftrightarrow\Be_4)$, $\Bdelta_1\leftrightarrow\Bdelta_2$ and $\Bdelta_3\leftrightarrow\Bdelta_4$.\footnote{Note that no permutation can mix the $\{\Be_i\}$ with the $\{\Bdelta_i\}$ in a Lie supergroup. In addition, in the non-compact part $\SU(2,2)$ there is no element of the Weyl group that is the analogue of $(\Be_1\leftrightarrow\Be_3,\Be_2\leftrightarrow\Be_4)$ in the compact sector.} The equality in \eqref{lmc} is understood to be mod $2\pi$.

What this discussion illustrates is that the undeformed and lambda model curves identify different ``special points", $z=1$ in the undeformed case, and the pair $z=\lambda^{\pm1/2}$ in the lambda model.

\subsection{Symplectic form}

The undeformed and lambda models share the same Lax connection, which implies that they have the same space of classical solutions (before imposing closed string boundary conditions). So a solution of the undeformed model can be mapped into a solution of the lambda model and vice-versa. However, not only do the theories have different constraints arising from the close string boundary conditions but they also have different symplectic forms and Poisson brackets. The Poisson brackets are described in Appendix \ref{a4}. 

We can write the symplectic form in terms of the wave function and, in particular, in terms of the one forms on the phase space $\Psi^{-1}\delta\Psi$ \cite{Appadu:2017xku}:\footnote{The phase space is identified with the space of classical solutions and $\Psi$ is defined on-shell and so depends implicitly on a point in the space of  classical solutions.}
\EQ{
\omega=\frac12\int_{-\pi}^\pi d\sigma\,\oint_{\cal C} du\,\STr\big\{\Psi^{-1}(z)\delta\Psi(z)\wedge\partial_1(\Psi^{-1}(z)\delta\Psi(z))\big\}\ .
\label{bol}
}
In the above, $du$ is a 1-form on the space of the spectral parameter $z$ 
that can be expressed in terms of the twist function,
\EQ{
du=\phi(z)\,\frac{dz}z\ ,\qquad \phi(z)=\frac k{8\pi}\cdot\frac{\lambda^2-\lambda^{-2}}{z^4-\lambda^2-\lambda^{-2}+z^{-4}}\ .
\label{twt}
} 
In the limit \eqref{liml}, we recover the twist function of the undeformed model \cite{Vicedo:2010qd}:
\EQ{
\phi(z)\Big|_\text{undeformed}=\frac{\kappa^2}{8\pi}\cdot\frac1{(z^2-z^{-2})^2}\ .
}
The twist function also appears in the definition of the Poisson brackets \eqref{gbb}. 
The contour in \eqref{bol} is chosen to encircle the poles of $\phi(z)$, i.e.~the special points $z=\lambda^{\pm1/2}$ and their images under the $\mathbb Z_4$ automorphism $z\to iz$.

The 1-form $du$ and the associated coordinate $u$ will play a key r\^ole for the CSC and subsequently for the QSC. In particular, the action coordinates on phase space are obtained from a generic CSC by integrating the meromorphic 1-form $p\,du$ involving the quasi-momenta:
\EQ{
I_{\cal C}=\oint_{\cal C}\frac{du}{2\pi i}\,p\ ,
}
where ${\cal C}$ are certain cycles on the CSC and $p$ corresponds to the quasi momenta viewed as a function over the spectral curve. The fact that the symplectic structure depends on the deformation parameter $\lambda$ means that the canonical action-angle variables of the lambda deformed theory are not the same as those in the undeformed theory. To complete the description of the dynamical system, the conjugate angle variables are identified with a point on the Jacobian torus of the CSC.

\subsection{Local conserved charges}

Integrable field theories like the present one, typically exhibit conserved charges that are both integrals of local functions of the fields and their derivatives as well as non-local charges. All these charges are encoded in the monodromy. In particular, in this section the focus will be on the local conserved charges. One of the motivations in constructing the local conserved charges is that they include the energy and momentum in the gauge fixed worldsheet theory. 

The solution of the Virasoro constraints imply that the Lax connection $\LAX_\mu(z)$ has poles at $z=0$ and $z=\infty$ proportional to an element which is conjugate to the Cartan element $\Lambda$. Choosing $z=\infty$, the argument starts by establishing that around the pole of the Lax connection at $\infty$, one can construct a gauge transformation $\Phi^{(\infty)}(z)=\sum_{n=0}^\infty \Phi^{(\infty)}_nz^{-n}$ order-by-order in $z^{-1}$ \cite{BBT},
\EQ{
\LAX_\mu^{(\infty)}(z)=\Phi^{(\infty)-1}(z)\partial_\mu\Phi^{(\infty)}(z)+\Phi^{(\infty)-1}(z)\LAX_\mu(z)\Phi^{(\infty)}(z)\ ,
\label{gt8}
}
such that $\LAX^{(\infty)}_1(z)$ commutes with its leading term proportional to $\Lambda$. This means that $\LAX^{(\infty)}_1(z)$ is block diagonal in index subsets $\{\hat 1,\hat 2,\tilde 1,\tilde 2\}$ and  $\{\hat 3,\hat 4,\tilde 3,\tilde 4\}$, corresponding to a pair of $\mathfrak{su}(2|2)$ subalgebras of $\mathfrak{psu}(2,2|4)$. Importantly, this can be done with $\Phi$ and $\LAX^{(\infty)}_1$ being a local function of the fields and their derivatives: there are no integrals. It follows that the gauge transformed monodromy
\EQ{
T^{(\infty)}(z)=\Phi^{(\infty)}(\pi;z)T(z)\Phi^{(\infty)-1}(-\pi;z)
}
is also block diagonal. Note that the boundary conditions, ensure that $\Phi^{(\infty)}(\pi;z)=\Phi^{(\infty)}(-\pi;z)$.

We can repeat the analysis expanding around $z=0$ constructing a gauge transformation $\Phi^{(0)}(z)$
\EQ{
\LAX_\mu^{(0)}(z)=\Phi^{(0)-1}(z)\partial_\mu\Phi^{(0)}(z)+\Phi^{(0)-1}(z)\LAX_\mu(z)\Phi^{(0)}(z)\ ,
\label{gf8}
}
where the right-hand side commutes with $\Lambda$.  It follows that the gauge transformed monodromy
\EQ{
T^{(0)}(z)=\Phi^{(0)}(\pi;z)T(z)\Phi^{(0)-1}(-\pi;z)
}
is also block diagonal in the same sense as $T^{(\infty)}(z)$.

Following the argument in \cite{Beisert:2005bm}, let us take one of the blocks, either $\{\hat 1,\hat 2,\tilde 1,\tilde 2\}$ or  $\{\hat 3,\hat 4,\tilde 3,\tilde 4\}$, indicated by the subscript $s\in\{1,2\}$. The gauge transformed monodromy of a particular block is given by
\EQ{
T^{(\infty)}_{(s)}(z)=\overset{\longleftarrow}{\text{Pexp}}\left[-\int_{-\pi}^{\pi}\, d\sigma\,\LAX^{(\infty)}_{1,(s)}(\sigma;z)\right]\ .
}
Taking the superdeterminant gives a quantity that can be written as an exponential integral but without path ordering. Hence, for the monodromy this yields a generator of local conserved quantities $q_{(s)}(z)$:
\EQ{
\text{sdet}\,T^{(\infty)}_{(s)}(z)=\exp[-i q_{(s)}(z)]\ ,\qquad
 q_{(s)}(z)=-i\int_{-\pi}^{\pi}\, d\sigma\, \STr\LAX^{(\infty)}_{1,(s)}(\sigma;z)\ .
}
In fact, there is only one independent generator $q_{(1)}(z)+q_{(2)}(z)=0$ since $\text{sdet}\, T(z)=\text{sdet}\,T^{(\infty)}(z)=\text{sdet}\,T_{(1)}^{(\infty)}(z)\text{sdet}\,T_{(2)}^{(\infty)}(z)=1$. We can express the $q_{(s)}(z)$ through the quasi-momenta
\EQ{
q_{(1)}(z)&=\hat p_1(z)+\hat p_2(z)-\tilde p_1(z)-\tilde p_2(z)\ ,\\ q_{(2)}(z)&=\hat p_3(z)+\hat p_4(z)-\tilde p_3(z)-\tilde p_4(z)\ .
}
The overall supertraceless combination is
\EQ{
\mathfrak Q(z)=q_{(1)}(z)-q_{(2)}(z)=(\Bxi_1-\Bxi_2)\cdot\Bp(z)\ .
\label{wes}
}
The charge generator $\mathfrak Q(z)$ corresponds to the current
\EQ{
\EuScript J^\mu(z)=2\epsilon^{\mu\nu}\STr\big\{\Lambda\LAX^{(\infty)}_{\nu}(z)\big\}\ ,
\label{ker}
}
whose conservation follows directly from the gauge transformed Lax equation \eqref{leq2}  from which it follows
\EQ{
\partial_\mu\EuScript J^\mu(z)=2\STr\big\{\Lambda[\partial_0+\LAX_0^{(\infty)}(z),\partial_1+\LAX_1^{(\infty)}(z)]\big\}=0\ ,
}
since $[\Lambda,\LAX^{(\infty)}_\mu(z)]=0$ on shell. 

A similar story plays out for the expansion around $z=0$. It is important that the Lax connection is analytic in the spectral parameter with poles at $z=0$ and $z=\infty$ and so $\mathfrak Q(z)$ can be obtained via an expansion around either $z=0$ or around $z=\infty$.

In appendix \ref{a4}, we argue that one of these charges, in the gauge fixed theory, is identified with the physical Hamiltonian and momentum on the worldsheet. We should emphasize that it is only in the gauge-fixed theory that it makes sense to define an energy: the Hamiltonian vanishes before gauge fixing. This identifies the energy and momentum of a configuration as
\EQ{
E&=(\lambda-\lambda^{-1})\big(\mathfrak Q(\lambda^{1/2})+\mathfrak Q(\lambda^{-1/2})\big)\\&=(\lambda-\lambda^{-1})(\Bxi_2-\Bxi_1)\cdot\big(\Bp(\lambda^{1/2})+\Bp(\lambda^{-1/2})\big)\ ,
\label{suc1}
}
and
\EQ{
P&=(\lambda+\lambda^{-1})\big(\mathfrak Q(\lambda^{1/2})-\mathfrak Q(\lambda^{-1/2})\big)\\&=(\lambda+\lambda^{-1})(\Bxi_2-\Bxi_1)\cdot\big(\Bp(\lambda^{1/2})-\Bp(\lambda^{-1/2})\big)\ .
\label{suc2}
}
Notice that the level matching condition \eqref{lmc} (with the identity permutation) in the lambda string implies the vanishing the momentum, up to some winding numbers, just as it does in the undeformed case.

\section{Classical Spectral Curve}\label{s4}

In this section, we construct an integral representation of the CSC of the lambda model. Since the two models share the same Lax connection and wave function, some of the discussion is similar to that of the undeformed model. In particular, we will follow quite closely the approach of Beisert et al.~\cite{Beisert:2005bm}, although we will write the curve in terms of Lie superalgebra roots which makes the subtle form of the curve more transparent. 

We have already remarked that the curves of the original sigma model and its associated deformed lambda model are the same. However, the associated meromorphic 1-forms $p\,du$ are different as are the physical conditions imposed by closed string boundary conditions. The 1-form $p\,du$ is, as we have described, determined by the symplectic structure of the theory and, in turn, it determines the action variables of the integrable system. Consequently, the construction of the curve of the lambda model has some differences compared with its undeformed cousin. Another difference is that the important conserved charge data encoded in the curve is associated to the behaviour of the quasi-momenta $\Bp$ at the pair of special points $z^4=\lambda^{\pm2}$, or $u=\pm\infty$, in the lambda model whereas in the original undeformed model these merge into a single point $z=1$.

\subsection{Base curve}

\begin{figure}[ht]
\begin{center}
\begin{tikzpicture}[scale=0.7]
\draw[draw=none,pattern=north west lines,opacity=0.4] (2,3.5) -- (4,3.5) -- (4,-0.5) -- (2,-0.5);
\draw[] (0,4) ellipse (0.25cm and 1cm); 
\draw[] (6,4) ellipse (0.25cm and 1cm); 
\draw[] (0,0) ellipse (0.25cm and 1cm); 
\draw[] (6,0) ellipse (0.25cm and 1cm); 
\draw[-] (0,5) -- (6,5);
\draw[-] (0,3) -- (6,3);
\draw[-] (0,1) -- (6,1);
\draw[-] (0,-1) -- (6,-1);
\begin{scope}[xscale=0.25]
\draw[very thick,->] (4,4) arc (-180:175:1);
\end{scope}
\node at (1,5.5) {\small $u\sim u+ik$};
\draw[thick,->] (0,4) -- (-0.5,4);
\draw[thick,->] (6,4) -- (6.5,4);
\draw[thick,->] (0,0) -- (-0.5,0);
\draw[thick,->] (6,0) -- (6.5,0);
\node at (-2,4.3) {\footnotesize$u=-\infty$};
\node at (8,4.3) {\footnotesize$u=\infty$};
\draw[thick,->] (6,4) -- (6.5,4);
\node at (-2,3.7) {\footnotesize$x=-1/\xi$};
\node at (8,3.7) {\footnotesize$x=\infty$};
\node at (-2,0.3) {\footnotesize$u=-\infty$};
\node at (8,0.3) {\footnotesize$ u=\infty$};
\draw[thick,->] (6,4) -- (6.5,4);
\node at (-2,-0.3) {\footnotesize$x=-\xi$};
\node at (8,-0.3) {\footnotesize$x=0$};
\draw[thick,decorate,
decoration={snake,amplitude=1pt, segment length=3pt}] (2,3.5) -- (4,3.5);
\filldraw[black] (2,3.5) circle (0.04cm);
\filldraw[black] (4,3.5) circle (0.04cm);
\node at (1.3,-0.65) {\footnotesize$-u_0$};
\node at (4.5,-0.7) {\footnotesize$u_0$};
\node (b1) at (1.9,-1.6) {\footnotesize$x=1$};
\node (b2) at (4.1,-1.6) {\footnotesize$x=-1$};
\draw[->] (b1) -- (2,-0.7);
\draw[->] (b2) -- (4,-0.7);
\node at (5,4.5) {\footnotesize$\Sigma^{(+)}$};
\node at (5,0.5) {\footnotesize$\Sigma^{(-)}$};
\draw[thick,decorate,
decoration={snake,amplitude=1pt, segment length=3pt}] (2,-0.5) -- (4,-0.5);
\filldraw[black] (2,-0.5) circle (0.04cm);
\filldraw[black] (4,-0.5) circle (0.04cm);
\draw[thick,dashed] (2,3.5) -- (2,-0.5);
\draw[thick,dashed] (4,3.5) -- (4,-0.5);
\end{tikzpicture}
\caption{\footnotesize  The underlying surface $\Sigma$, the double cover of the cylinder $u\sim u+ik$,  with its two sheets $\Sigma^{(\pm)}$ joined by the short cut $[-u_0,u_0]$.}
\label{fig2} 
\end{center}
\end{figure}
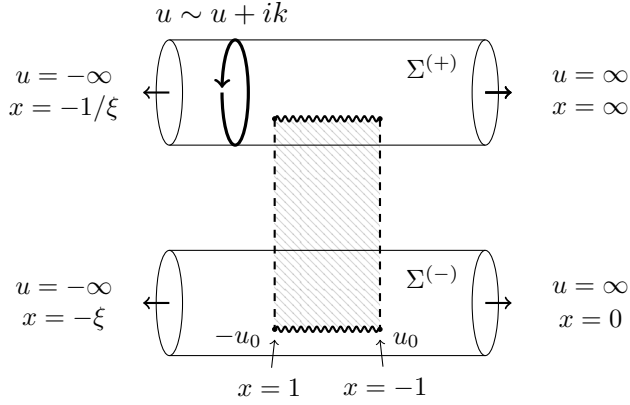

The CSC is the algebraic curve defined in \eqref{d33}. We can think of it as a branched covering over the base defined by complex plane of the spectral parameter $z$. In fact, in the discussion, it is useful to introduce two alternative spectral parameters, $u$ and $x$. The first alternative, $u$ is obtained by integrating the 1-form $du$ in \eqref{twt}, 
\EQ{
z^4=\frac{\sinh[\pi(u+u_0)/k]}{\sinh[\pi(u-u_0)/k]}\ ,
\label{vvz}
}
where we define the constant $u_0$ via
\EQ{
e^{\pi u_0/k}=\frac1\lambda\ .
}
So we can think of the $z$ plane as 4-fold cover of the complex $u$ cylinder $u\sim u+ik$. 
The quasi-momenta take values in the even part of the superalgebra and so it is natural to focus on $z^2$ rather than $z$ itself. In that case, we can think of the $z^2$ plane as being two copies of the $u$ cylinder that we denote as $\Sigma^{(\pm)}$, joined by the branch cut $[-u_0,u_0]$, as shown in Figure \ref{fig2}. The value of $z^2$ on $\Sigma^{(\pm)}$ are related by $z^2\to -z^2$, consistent with the $\mathbb Z_4$ structure of the underlying twisted affine algebra. 

The other spectral parameter is the ``Zhukovsky" parameter $x$ which is related to $z^2$ by a simple transformation
\EQ{
z^2=\frac1\lambda\cdot\frac{x+1}{x-1}\ .
\label{xzz}
}
So the two branches $\Sigma^{(\pm)}$ are related by $x\to1/x$, i.e.~$z^2\to-z^2$, which is generated by the $\mathbb Z_4$ automorphism on the spectral parameter $z\to iz$.

The relationship between $x$ and $u$ is then determined to be of the form
\EQ{
x+\frac1x=\Big(\frac1\xi-\xi\Big)e^{2\pi u/k}-\frac1\xi-\xi\ ,
\label{fv2}
}
where we have introduced a constant $0\leq \xi\leq 1$ defined via
\EQ{
\xi=\frac{1-\lambda^2}{1+\lambda^2}\ .
}
This is a generalization of the Zhukovsky map $x+1/x=u/g$ of the undeformed model; indeed, in the limit $k\to\infty$, $\lambda\to1$, with $g=k(1-\lambda)/2\pi$ fixed, it reduces precisely to this. This mapping first appeared in the context of the S-matrix theory of the excitations of the gauge-fixed lambda model \cite{Hoare:2011wr}. Note that the branch points $u=\pm u_0$, correspond to $x=\pm1$. A given point $u$ on the cylinder has images $x$ and $1/x$, with $|x|\gtrless1$, on $\Sigma^{(\pm)}$, respectively.

\subsection{Special points}
 
A particularly important r\^ole is played by the points at infinity in the $u$ parameterization, $u=\pm\infty$. These correspond to $z^4=\lambda^{\pm2}$ or 
$x=\infty$ and $x=-1/\xi$, respectively, on $\Sigma^{(+)}$, and $x=0$ and $x=-\xi$, respectively, on $\Sigma^{(-)}$. These points are related directly to the conserved charges of the two Kac-Moody symmetries of the theory, associated to the currents \eqref{kmc}, and these charges include the world-sheet energy in the gauge-fixed lambda model \cite{Appadu:2017xku}.

Note that these special values are at the poles of the twist function defined in \eqref{twt}. Before discussing the special points in the lambda model, let us first remind ourselves what happens in the undeformed model. In this case, the mapping \eqref{vvz} becomes
\EQ{
z^4=\frac{u+2g}{u-2g}\ ,
}
and \eqref{fv2} reduces to the usual Zhukovsky map $x+1/x=u/g$. There is now a single point at infinity, or ``special" point,  $u=\infty$, i.e.~$z^4=1$. If we expand around $z=1$, the Lax connection in the undeformed model has the asymptotic form
\EQ{
\LAX_\pm=f^{-1}\partial_\pm f+\frac gu\big((f^{-1}\partial_\pm f)^{(1)}\mp 2(f^{-1}\partial_\pm f)^{(2)}-(f^{-1}\partial_\pm f)^{(3)}\big)+{\mathscr O}(u^{-2})\ .
}
Making a gauge transformation $\LAX_\pm\to\LAX'_\pm=f(\partial_\pm+\LAX_\pm)f^{-1}$,
\EQ{
\LAX'_\pm=\frac guf\big((f^{-1}\partial_\pm f)^{(1)}\mp 2(f^{-1}\partial_\pm f)^{(2)}-(f^{-1}\partial_\pm f)^{(3)}\big)f^{-1}+\mathscr O(u^{-2})\ ,
}
and so the zero curvature condition at order $1/u$ has the form of a conservation equation for the Noether current
\EQ{
J^L_\pm= f\big(\mp\frac12 (f^{-1}\partial_\pm f)^{(1)}+(f^{-1}\partial_\pm f)^{(2)}\pm\frac12(f^{-1}\partial_\pm f)^{(3)}\big)f^{-1}\ ,
}
for the global $\text{PSU}(2,2|4)$ left symmetry $f\to Uf$ of the undeformed model. So $\LAX_\pm$ is related to the components of the current $\pm J^L_\pm$ by a gauge transformation by $f$. But this gauge transformation does not affect the spectrum of the monodromy, which is invariant provided that $f$ satisfies periodic boundary conditions. Hence, this means that as $u\to\infty$, the quasi-momenta have the asymptotic behaviour $p_A(u)=2gQ_A/u+\mathscr O(u^{-2})$. where $Q_A$ are the Noether charges for the global $\text{PSU}(2,2|4)$ left symmetry. In the gauge-fixed worldsheet theory, these charges include the energy of the states.

Now let us turn to the lambda model. Here, there are two special points, $z=\lambda^{\pm1/2}$ (up to automorphism), or $u=\mp\infty$. 
At these points, the spatial component of the Lax connection becomes proportional to the Kac-Moody currents \eqref{kmc}:
\EQ{
\LAX_1(u=\mp\infty)=\mp\frac{2\pi}k \JJ_\pm\ ,
\label{kmm}
}
where
\EQ{
\JJ_+&=-\frac k{2\pi}\big(\CF^{-1}\partial_+\CF+\CF^{-1}A_+\CF-A_-\big)\ ,\\
\JJ_-&=\frac k{2\pi}\big(\partial_-\CF\CF^{-1}-\CF A_-\CF^{-1}+A_+\big)\ .
}
So at the special points, the quasi-momenta are the non-abelian charges associated to the Kac-Moody symmetry. 

Following the notation of \cite{Gromov:2014caa} (and the excellent review \cite{Gromov:2017blm}), we will write the quasi-momenta $p_A=(\tilde p_i,\hat p_i)$, for $i\in\{1,2,3,4\}$ associated to the $\mathfrak{su}(4)$ and $\mathfrak{su}(2,2)$ bosonic sub-algebras of $\mathfrak{psu}(2,2|4)$, respectively. 
The charges are extracted from the quasi-momenta at the special points $z=\lambda^{\mp1/2}$, i.e.~$u=\pm\infty$. Using this notation, we will define the charges as
\EQ{
\tilde p_i(u=\pm\infty)=\frac{2\pi}k\lambda^{(\pm)}_i\ ,\qquad \hat p_i(u=\pm\infty)=-\frac{2\pi}k\nu^{(\pm)}_i\ .
\label{uiu}
}
where the $(\lambda_i^{(\pm)},\nu_i^{(\pm)})$ are understood to be defined modulo $k$ since only the eigenvalues of the monodromy $\exp(ip_A)$ are physically significant.

The charges define a pair of weight vectors
\EQ{
\Bw^{(\pm)}=\sum_i\big(\lambda_i^{(\pm)}\Be_i+\nu_i^{(\pm)}\Bdelta_i\big)=
\frac k{2\pi}\Bp(z=\lambda^{\mp1/2})\ ,
\label{dww}
}
which are related by the boundary conditions. With closed string boundary conditions, the level matching conditions \eqref{lmc} state that
\EQ{
(\lambda_i,\nu_i)\equiv (\lambda_i^{(+)},\nu_i^{(+)}) = {\mathbb P}(\lambda_i^{(-)},\nu_i^{(-)})\ .
\label{cgg}
}
The equality in here is understood to be mod $k$.
The CSC can be constructed with arbitrary weights $\Bomega^{(\pm)}$ and the level matching condition can be imposed ex post facto. 

The relation \eqref{dww} deserves some comment. Weight vectors should satisfy the vanishing of the central charge condition \eqref{ccc}. This condition is actually implied by the unimodularity condition on the quasi-momenta
\EQ{
\sum_{i=1}^4\hat p_i(z)=\sum_{i=1}^4\tilde p_i(z)\ .
\label{unii}
}
This is to be expected because the charges, or weights, and the quasi-momentum are dual under the supertrace.\footnote{A $z$-dependent charge is an object $\mathfrak Q_q(z)=\STr(q\, p(z))$ 
defined in terms of an algebra element $q$ and the quasi-momentum $p(z)=\Bp(z)\cdot\BH$. The charge of the central element $q=\mathfrak C=i 1_8$ vanishes precisely because of the unimodularity constraint $\STr(p(z))=0$.
}

The undeformed model has a global $\text{PSU}(2,2|4)$ symmetry whereas we expect the lambda model to have a quantum group symmetry $\PSU_q(2,2|4)$, $q^{k}=-1$.\footnote{This expectation is based on the symmetries of the bosonic lambda models, e.g.~\cite{Appadu:2017fff} and also those of integrable deformations of WZW \cite{Ahn:1990gn}. The quantum group is a deformation of the universal enveloping of $\mathfrak{psu}(2,2|4)$, a Drinfeld-Jimbo algebra, e.g.~see the books \cite{KS,CP}. The deformation parameter $q$ is a root-of-unity and in that case, the quantum supergroup has additional central elements. We are interested in the {\it restricted\/} quantum group obtained by quotienting by these additional central elements. So by $\PSU_q(2,2|4)$ we mean the restricted quantum supergroup.} Both the group and the quantum group share the same Cartan generators and so the structure of the charges is the same. In particular, we can identify the weights of the superalgebra with those charges, as in the undeformed model, following the conventions of \cite{Gromov:2013pga}, as
\EQ{
\lambda_i=\frac12\begin{cases}J_1+J_2-J_3&\\ J_1-J_2+J_3 &\\ -J_1+J_2+J_3 &\\ -J_1-J_2-J_3\end{cases},\qquad
\nu_i=\frac12\begin{cases}-\Delta+S_1+S_2 &\\ -\Delta -S_1-S_2 &\\ \Delta+S_1-S_2 &\\ \Delta-S_1+S_2 \end{cases}\ ,
\label{dat}
} 
where $J_{1,2,3}$ and $S_{1,2}$ are associated to the quantum group versions of the compact subalgebras, $\mathfrak{su}(4)$ and $\mathfrak{su}(2)\oplus\mathfrak{su}(2)\subset\mathfrak{su}(2,2)$, respectively, and so are half integers, while $\Delta$ is associated to the non-compact part of the subalgebra $\mathfrak{su}(2,2)$ and is not quantized. Note, in the classical theory, valid for large $k$, the charges scale like $k$.

The energy of the gauge-fixed theory are identified with the following combination of charges \eqref{suc1}, \eqref{suc2}. Choosing the trivial permutation in \eqref{cgg}, this gives
\EQ{
E=\frac{4\pi}k(\lambda^{-1}-\lambda)\big(\Delta-J_1\big)\ ,
}
just as in the $\text{AdS}_5{\times} S^5$ string, while the level matching condition requires the vanishing of the momentum (mod $k$).

\subsection{Engineering the curve}

Now we turn to the CSC itself. Some of the details are similar to the undeformed case described at length \cite{Beisert:2005bm} although here we will keep the relation with the root system of the Lie superalgebra more transparent. The CSC is the branched 8-sheeted covering of the base $\Sigma$ defined by the eigenvalues $\exp[ip_A(x)]$ which we associate to a weight vector $\Bp(x)$ as in \eqref{goo}, although here we find it more useful to 
think of it as a function of the spectral parameter $x$ which covers both sheets $\Sigma^{(\pm)}$. The quasi-momenta has various properties that we list below. 

\vspace{0.2cm}
\noindent{\bf (i)} The $\mathbb Z_4$ action of the automorphism acts on the base as $z^2\mapsto-z^2$, or $x\mapsto1/x$, and implies the following action on the quasi-momenta
\EQ{
\Bp(1/x) =\sigma_-(\Bp(x))+2\pi m_1\Bxi_1+2\pi m_2\Bxi_2\ .
\label{ccf}
}
This means that the symmetry exchanges the values of the quasi-momenta on the second sheet $\Sigma^{(-)}$ with those on the physical sheet $\Sigma^{(+)}$ along with a permutation of the index. We can think of the integers $m_j$ as winding numbers.

\vspace{0.1cm}\noindent{\bf(ii)} The fact that we are working in $\text{PSU}(2,2|4)$ rather than $\text{U}(2,2|4)$ has the following implications. Firstly, unimodularity, the ``S'', means that the quasi-momenta are subject to the constraint \eqref{unii}. Projectivity, the ``P'', means that shifts by a central element are not physical:
\EQ{
\Bp(x)\thicksim\Bp(x)+\zeta(x)\sum_{i=1}^4\big(-\Be_i+\Bdelta_i\big)\ ,
\label{vvr4}\
}
for  an arbitrary function $\zeta(x)$, with $\zeta(1/x)=-\zeta(x)$ so that the shift has the correct grade in $\hat\mf$.

\vspace{0.1cm}\noindent{\bf(iii)}  The Virasoro constraints \eqref{epk}, specify the poles of the quasi-momenta at $z\to 0$ and $z\to\infty$, i.e.~$x\to\mp1$, respectively. Since the Lax operator has the poles at $z\to0,\infty$ \eqref{jla},
\EQ{
\LAX_1\longrightarrow \pm\mu_\pm z^{\pm2} V_\pm\Lambda V_\pm^{-1}+\cdots\ ,
} 
the quasi-momenta must have corresponding poles at $x=\pm1$ which we can incorporate by writing
\EQ{
\Bp(x)=\sum_{s=\pm}{\cal P}_s(x)\sum_{l=1}^2a_{l,s}\Bxi_l+\cdots\ ,
\label{nnv}
}
for constants $a_{s,l}$ subject to the constraints 
\EQ{
\big(\xi a_{1,+}\pm a_{1,-}\big)^2=\big(\xi a_{2,+}\pm a_{2,-}\big)^2\ .
\label{fca}
} 
In \eqref{nnv}, the ellipsis is regular at $x=\pm1$ and we have defined the functions
\EQ{
{\cal P}_\pm(x)=\frac{z(x)^2\pm z(x)^{-2}}{\lambda^{-1}\pm\lambda}=\frac{x^2+2\xi^{\pm1} x+1}{x^2-1}\ .
\label{dwp}
}
Notice that ${\cal P}_\pm(x)$ are consistent with the symmetries of the curve
\EQ{
{\cal P}_\pm(1/x)=-{\cal P}_\pm(x)\  .
}

\vspace{0.1cm}\noindent{\bf(iv)}  There are square root branch cuts associated to even roots of the superalgebra $\Balpha=\Be_i-\Be_j$ (or $\Bdelta_i-\Bdelta_j$)  that join a pair of sheets labelled by $\Be_i$ and $\Be_j$ (or $\Bdelta_i$ and $\Bdelta_j$) on the sheet  $\Sigma^{(+)}$ of the base. Each such cut comes in a pair related by the symmetry $x\mapsto1/x$, with a  permutation of the sheets according to the action of the automorphism in \eqref{zz13}. More specifically, the square root branch cuts are in the eigenvalues $\exp[ip_A(x)]$ rather than the quasi-momenta themselves and this means that the latter can jump by multiples of $2\pi$ on crossing a cut. We can express this by saying that the averages of the quasi-momenta across a cut ${\cal C}$ that joins the two sheets, must differ by an integer multiple of $2\pi$,
\EQ{
\Balpha\cdot\Bpslash(x)=2\pi n\ ,\quad x\in{\cal C}\ ,\quad n\in\mathbb Z\ .
\label{fll}
}
where $\Balpha$ is the associated even root. In the above, we have defined $\Bpslash(x)=(\Bp(x_+)+\Bp(x_-))/2$, where  $x_\pm$ lie infinitely close, but on either  side, of a point $x$ on the cut. The {\it filling\/} $K$ of a cut associated to a root $\Balpha$ is defined using the symplectic 1-form $du$ as 
\EQ{
K\Balpha=-\oint_{{\cal A}}\frac{du}{2\pi i}\,\Bp(x(u))\ ,
\label{df3}
}
where ${\cal A}$ in a contour that encircles the cut ${\cal C}$ on the sheet $\Be_i$ in a positive sense (or $\Be_j$ in a negative sense) and similarly for $\Bdelta_i-\Bdelta_j$.

So each cut has a filling $K$ and an integer $n$. This data is the curved space analogue of the amplitude and the mode number of a string in flat space.

\vspace{0.1cm}\noindent{\bf(v)}  Fermionic excitations are associated to fermionic roots $\Balpha=\Be_i-\Bdelta_j$, but instead of being associated to cuts they correspond to isolated poles with equal residues on each of the sheets $\Be_i$ and $\Bdelta_j$ in the pair. The poles also come in pairs associated to the transformation $x\mapsto1/x$. The residue should be understood as the product of two Grassmann numbers. In this case, we define $\Bpslash(x)$ as the regular part at the pole and then both \eqref{fll} and \eqref{df3} are still valid.

\vspace{0.1cm}\noindent{\bf(vi)}  In the lambda model, the quasi-momenta at the special points determine the charges of the configuration, a pair of weight vectors $\Bomega^{(\pm)}$ in \eqref{dww}.  This data is determined by the fillings of the cuts and the integer data in a way that we will uncover. These quantum numbers include, in the gauge-fixed worldsheet theory, the energy of a configuration. 

\vspace{0.2cm}
The CSC can be thought of as the solution of a Riemann-Hilbert problem for the quasi-momenta $\Bp$ defined over the base curve $\Sigma$ with the analytic properties outlined above. The solution of this Riemann-Hilbert problem can be formulated in terms of  seven independent ``densities", $\rho_r(x)$, $r=1,2,\ldots,7$, that can naturally be associated to a set of simple roots $\{\Balpha_r\}$ of the superalgebra associated to some ordering of the sheets $\Be_A$. A cut in the quasi-momentum associated to an even root $\Balpha=\sum_{r\in S}\Balpha_r$ for some $S\subset\{1,2,\ldots,7\}$, corresponds to an equal contribution to the densities of $\rho_r(x)$, for $r\in S$, of a function with support along the image of the cut (an open contour). If the root is odd, then the function has delta function support.

We will need two kinds of resolvents defined in terms of the densities. The first is defined solely in terms of the Zhukovsky spectral parameter $x$ and a suitable kernel $g(x,y)$:
\EQ{
G_r(x)=\frac{\pi}k\int dy\,\rho_r(y)g(x,y)\ .
\label{prs}
}
The basic property of the resolvent is that it is an analytic function with cuts along the support of the density for even roots  excitations and poles for fermionic excitations. The basic requirement, for the cuts, is that the discontinuity across a cut gives the density, along with a Jacobian:
\EQ{
\text{Disc}\,G_r(x)=2\pi i\cdot\frac{dx}{du}\cdot\rho_r(x)=\frac{4\pi^2i}k\cdot\frac{x+x^{-1}+\xi+\xi^{-1}}{1-x^{-2}}\cdot\rho_r(x)\ ,\qquad x\in{\cal C}_r\ .
\label{dcn}
}
Hence, the kernel $g(x,y)$ must have a pole at $x=y$ with residue
\EQ{
g(x,y)=\frac{k}\pi\cdot \frac{dx}{du}\cdot \frac1{x-y}+\cdots\ .
}
This does not completely determine $g(x,y)$ so there is some freedom here that is ultimately just a matter of convention, but we make the following choice, based on hindsight afforded by the semi-classical limit of the QSC that we discuss in \cite{qsc},
\EQ{
g(x,y)=\frac{(1+\xi y)(2xy+\xi y+\xi x)}{\xi(y^2-1)}\cdot\frac{1}{x-y}\ .
\label{uff}
}

Note that the defining property \eqref{dcn}, ensures that an integral of the 1-form $G_r(x)du/2\pi i$ around a cycle ${\cal A}$ that encloses the cut ${\cal C}_{r,p}$, where $p$ labels the cut, or surrounds a pole, gives the integral of the density, or the ``filling".
To see this, one shrinks ${\cal A}$ onto the cut and then expands at each point of the support $x=y$:
\EQ{
\oint_{{\cal A}}\frac{du}{2\pi i}\,& G_r(x(u))\\ &=\frac{2\pi}k\int_{{\cal C}_{r,p}} dy\,\rho_r(y)\oint_{y}\frac{du}{2\pi i}
\cdot\frac{x(u)+x(u)^{-1}+\xi+\xi^{-1}}{1-x(u)^{-2}}\cdot\frac1{x(u)-y}\\ &=\int_{{\cal C}_{r,p}} dy\,\rho_r(y)\oint_y\frac{dx}{2\pi i}
\cdot\frac1{x-y}\\ &=\int_{{\cal C}_{r,p}} dy\,\rho_r(y)=K_{r,p}\ .
\label{fpp}
}
For later we will define the total filling of each density function $\rho_r(x)$
\EQ{
K_r=\sum_pK_{r,p}\ .
}

For an odd root, say $\Balpha=\Bdelta_i-\Be_j$, the cuts are replaced by isolated poles which we can also express via kernel $g(x,y)$. Near $x=x_{p}$, the position of the pole,
\EQ{
\hat p_i(x)=-\kappa_p g(x,x_p)+\cdots\ ,\qquad \tilde p_j(x)=-\kappa_p g(x,x_p)+\cdots\ .
}
The residues $\kappa_p$ are products of two Grassmann numbers. The resolvent $G_r(x)$ for an odd simple root, is defined as a sum over the poles
\EQ{
G_r(x)=\sum_p\kappa_{r,p}g(x,x_{r,p})\ .
}

With these preliminaries, we can make an ansatz for the quasi-momenta that incorporates all the constraints and symmetries:
\EQ{
\Bp(x)=-\sum_{r=1}^7\big(\Balpha_rG_r(x)+\sigma_-(\Balpha_r)G_r(1/x)
\big)+\sum_{s=\pm}{\cal P}_s(x)\sum_{l=1}^2a_{l,s}\Bxi_l+\Bphi\ ,
\label{xpi}
}
where $\Bphi$ is the constant vector 
\EQ{
\Bphi=\pi m_1\Bxi_1+\pi m_2\Bxi_2\ .
\label{dph1}
}
Note that we could add a trivial shift to $\Bphi$ in the form of a constant element of the Cartan subalgebra of $f^{(0)}$. The engineered curve above depends implicitly on the densities $\rho_r(x)$ as well as the constants $a_{l,s}$ constrained via \eqref{fca} and $m_1$ and $m_2$.

Let us write the curve in a different way which will facilitate a simpler comparison with the semi-classical limit of the Bethe Ansatz equations in the companion work \cite{qsc} (see also section \ref{s6}), by introducing an inversion-symmetric resolvent naturally expressed in terms of the coordinate $u$:
\EQ{
H_r(x)&=G_r(x)+G_r(1/x)-\frac12\big(G_r(\infty)+G_r(-1/\xi)\big)\\ &=\frac\pi {k}\int dy\,\rho_r(y)\coth\big[\pi(u(x)-u(y))/k\big]\ .
\label{hde}
}
Using the above, we can write the curve as
\EQ{
\Bp(x)=-\sum_{r=1}^7\big(\Balpha_rH_r(x)+(\sigma_--1)\Balpha_rG_r(1/x)\big)+\sum_{s=\pm}{\cal P}_s(x)\sum_{l=1}^2a_{l,s}\Bxi_l+\Bphi'\ ,
\label{gut1}
}
where 
\EQ{
\Bphi'=\Bphi-\frac12\sum_{r=1}^7\Balpha_r\big(G_r(\infty)+G_r(-1/\xi)\big)\ .
}

We remark here, that $1-\sigma_-$ projects onto the component $\mf^{(2)}$ that lies in the Cartan subalgebra. This subspace is spanned by the two vectors $\Bxi_1$ and $\Bxi_2$ defined in \eqref{fll}, so we can write\footnote{This relation is valid up to a shift \eqref{vvr4}.}
\EQ{
\sum_{r=1}^7(\sigma_--1)\Balpha_rG_r(1/x)=\beta_1(1/x)\Bxi_1-\beta_2(1/x)\Bxi_2\ ,
}
where we have defined quantities, for $j=1,2$,
\EQ{
\beta_j(x)=\frac12\Bxi_j\cdot\sum_{r=1}^7\Balpha_rG_r(x)\ .
\label{dfb}
}
The final expression for the curve is
\EQ{
\Bp(x)=-\sum_{r=1}^7\Balpha_rH_r(x)-\beta_1(1/x)\Bxi_1+\beta_2(1/x)\Bxi_2+\sum_{s=\pm}{\cal P}_s(x)\sum_{l=1}^2a_{l,s}\Bxi_l+\Bphi'\ .
\label{rr6}
}

Hence, we can write the conditions that determine the curve \eqref{fll} as a set of integral equations for the densities
\EQ{
\sum_{s=1}^7\Balpha_r\cdot\Balpha_s\Hslash_s(x)+F_r(x)=2\pi n_{r,p}\ ,\qquad x\in{\cal C}_{r,p}\ ,\quad n_{r,p}\in\mathbb Z\ ,
\label{ccv2}
}
where the general expression for the ``driving terms" is
\EQ{
F_r(x)= \Balpha_r\cdot\big(\beta_1(1/x)\Bxi_1-\beta_2(1/x)\Bxi_2-\sum_{s=\pm}{\cal P}_s(x)\sum_{l=1}^2a_{l,s}\Bxi_l-\Bphi'\big)\ .
\label{ddr}
}
Notice that $F_r(x)$ is analytic along the cuts ${\cal C}_{r,p}$.
The way we have constructed the curve in terms of integral equations for densities is a characteristic way of solving a Riemann-Hilbert problem.

To summarize, we have defined an integral representation of the CSC in terms of the densities $\rho_r(x)$ which have to satisfy the auxiliary conditions \eqref{ccv2}.
It is a simple matter to extract the charges of the configuration by looking at the behaviour of the quasi-momenta at the special points \eqref{dat}:
\EQ{
\Bomega^{(\pm)}=\mp\frac12\sum_{r=1}^7K_r\Balpha_r\mp\frac k{2\pi}\big(\beta_1(0)\Bxi_1-\beta_2(0)\Bxi_2-\sum_{l=1}^2(a_{l,-}\pm a_{l,+})\Bxi_l\mp\Bphi'\big)\ .
}
In writing the above, we have used the fact that $G_r(0)=-G_r(-\xi)$ and so $\beta_r(0)=-\beta_r(-\xi)$.

We have defined the curve for general $\Bomega^{(\pm)}$ and for closed strings we need to impose the level matching condition \eqref{lmc}.

\subsection{A particular choice of simple roots}

In order to make contact with \cite{Beisert:2005bm}, let us consider one particular choice of simple roots, precisely those described by the KDV diagram in \eqref{bn1}. In that case, we can write the components of the quasi-momentum, up to a shift \eqref{vvr4}, explicitly as 
\EQ{
\hat p_i(x)&=\hat H_i(x)+\varepsilon_i\big(\beta_1(1/x)+a_{2,+}{\cal P}_+(x)+a_{2,-}{\cal P}_-(x)\big)+\hat\phi_i'\ ,\\[5pt]
\tilde p_i(x)&=\tilde H_i(x)+\varepsilon_i\big(\beta_2(1/x)-a_{1,+}{\cal P}_+(x)-a_{1,-}{\cal P}_-(x)\big)+\tilde\phi_i'\ ,
\label{qmm}
}
where we have defined $\varepsilon_{i}=(1,1,-1,-1)$ and 
\EQ{
\hat H_i=\begin{cases}H_1 &\\ H_4-H_3 &\\ H_5-H_4 &\\-H_7\end{cases},\qquad\tilde H_i=\begin{cases}H_1-H_2&\\ H_2-H_3&\\ H_5-H_6&\\H_6-H_7\end{cases}.
\label{udu}
}
In the above,
\EQ{
\beta_1(x)&=-\frac12(G_1(x)-G_3(x)-G_5(x)+G_7(x)+2G_4(x))\ ,\\
\beta_2(x)&=-\frac12(G_1(x)-G_3(x)-G_5(x)+G_7(x))\ .
}
The ``driving terms" $F_r(x)$ are, from \eqref{ddr},
\EQ{
F_{1}(x)&=G_4(1/x)+u_1(x)+u_2(x)\ ,\quad F_2(x)=0\ ,\\ 
F_{3}(x)&=-G_4(1/x)-u_1(x)-u_2(x)\ ,\\
F_4(x)&=G_1(1/x)-G_3(1/x)-G_5(1/x)+G_7(1/x)+2G_4(1/x)+2u_1(x)\ ,\\
F_{5}(x)&=-G_4(1/x)-u_1(x)-u_2(x)\ ,\\ F_6(x)&=0\ ,\quad
F_{7}(x)=G_4(1/x)+u_1(x)+u_2(x)\ ,
\label{drt}
}
where
\EQ{
u_l(x)=\sum_{s=\pm}a_{l,s}{\cal P}_s(x)\ .
}

In order to make contact with the gauge-fixed theory, we must impose conditions on the residues of the curve at $x=\pm1$; namely, $a_{1,+}=-a_{2,+}$ and $a_{l,-}=0$ in \eqref{gut1}.\footnote{In fact, this choice corresponds to the choice of reference solution \eqref{yer2}. It is also possible to choose $a_{1,+}=-a_{2,-}/\xi$ and $a_{1,-}=a_{2,+}=0$ which corresponds to the alternative reference solution \eqref{yer4}.} These conditions are consistent with the fact that in the gauge fixed theory described in detail in appendix \ref{a4}, has poles of form
\EQ{
\LAX_+=\mu z^2\Lambda+\cdots\ ,\qquad \LAX_-=-\mu z^{-2}\gamma^{-1}\Lambda\gamma+\cdots\ .
}
In the gauge-fixed theory, the energy of a configuration is given by \eqref{suc1},
\EQ{
E&=2(\lambda^{-1}-\lambda)(\Bxi_1-\Bxi_2)\cdot\Bphi'\\ &=2(\lambda^{-1}-\lambda)\Big((\Bxi_1-\Bxi_2)\cdot\Bphi+\beta_2(\infty)+\beta_2(-1/\xi)-\beta_1(\infty)-\beta_1(-1/\xi)\Big)\\
&=-8\pi(\lambda^{-1}-\lambda)(m_1+m_2)+\frac{4\pi}k(\lambda^{-1}-\lambda)\int dx\,\rho_4(x){\cal P}_-(x)\ .
}
The momentum is given by \eqref{suc2}
\EQ{
P&=\frac{4\pi}k(\lambda^{-1}+\lambda)\Big(K_4+\frac{2k}\pi\beta_1(0)-\frac{2k}\pi\beta_2(0)\Big)\\
&=\frac{4\pi}k(\lambda^{-1}+\lambda)\Big(K_4-\frac{2k}\pi G_4(0)\Big)\\ &=\frac{4\pi}k(\lambda^{-1}+\lambda)\int dx\,\rho_4(x){\cal P}_+(x)\ .
\label{eeg}
}
Note that it is the density $\rho_4(x)$ that contributes to the energy and momentum.

\subsection{More general bases}\label{s4.6}

It is interesting to construct the form of the CSC for four choices of simple roots, three additional ones to add to the choice in the last section. These are choices in \cite{Beisert:2005fw} that correspond to four different ordering of the Riemann sheets with the following four KDV diagrams:
\begin{equation*}
\begin{tikzpicture}[scale=0.7]
\begin{scope} [yshift=6cm,scale=0.7]
\node at (-7,0) {$\eta_1=1,\eta_2=1$};
\node at (1,1) {$\hat1$};
\node at (-1,1) {$\tilde1$};
\node at (5,1) {$\tilde2$};
\node at (3,1) {$\hat2$};
\node (a1) at (9,1) {$\hat3$};
\node (a2) at (7,1) {$\tilde3$};
\node at (13,1) {$\tilde4$};
\node at (11,1) {$\hat4$};
\draw (0.5,0) -- (1.5,0);
\draw (2.5,0) -- (3.5,0);
\draw (4.5,0) -- (5.5,0);
\draw (6.5,0) -- (7.5,0);
\draw (8.5,0) -- (9.5,0);
\draw (10.5,0) -- (11.5,0);
\draw (2,0) circle (0.5cm);
\draw (6,0) circle (0.5cm);
\draw (10,0) circle (0.5cm);
\begin{scope}[xshift=0cm]
\draw (0,0) circle (0.5cm);
\draw (0,0) circle (0.4cm);
\draw (0.285,0.285) -- (-0.285,-0.285);
\draw (0.285,-0.285) -- (-0.285,0.285);\end{scope}
\begin{scope}[xshift=4cm]
\draw (0,0) circle (0.5cm);
\draw (0,0) circle (0.4cm);
\draw (0.285,0.285) -- (-0.285,-0.285);
\draw (0.285,-0.285) -- (-0.285,0.285);
\end{scope}
\begin{scope}[xshift=8cm]
\draw (0,0) circle (0.5cm);
\draw (0.353,0.353) -- (-0.353,-0.353);
\draw (0.353,-0.353) -- (-0.353,0.353);
\end{scope}
\begin{scope}[xshift=12cm]
\draw (0,0) circle (0.5cm);
\draw (0.353,0.353) -- (-0.353,-0.353);
\draw (0.353,-0.353) -- (-0.353,0.353);
\end{scope}
\end{scope}
\begin{scope} [yshift=4cm,scale=0.7]
\node at (-7,0) {$\eta_1=-1,\eta_2=1$};
\node at (1,1) {$\hat1$};
\node at (-1,1) {$\tilde1$};
\node at (5,1) {$\tilde2$};
\node at (3,1) {$\hat2$};
\node (a1) at (7,1) {$\hat3$};
\node (a2) at (9,1) {$\tilde3$};
\node at (11,1) {$\tilde4$};
\node at (13,1) {$\hat4$};
\draw (0.5,0) -- (1.5,0);
\draw (2.5,0) -- (3.5,0);
\draw (4.5,0) -- (5.5,0);
\draw (6.5,0) -- (7.5,0);
\draw (8.5,0) -- (9.5,0);
\draw (10.5,0) -- (11.5,0);
\draw (2,0) circle (0.5cm);
\draw (10,0) circle (0.5cm);
\begin{scope}[xshift=0cm]
\draw (0,0) circle (0.5cm);
\draw (0,0) circle (0.4cm);
\draw (0.285,0.285) -- (-0.285,-0.285);
\draw (0.285,-0.285) -- (-0.285,0.285);
\end{scope}
\begin{scope}[xshift=4cm]
\draw (0,0) circle (0.5cm);
\draw (0,0) circle (0.4cm);
\draw (0.285,0.285) -- (-0.285,-0.285);
\draw (0.285,-0.285) -- (-0.285,0.285);
\end{scope}
\begin{scope}[xshift=6cm]
\draw (0,0) circle (0.5cm);
\draw (0.353,0.353) -- (-0.353,-0.353);
\draw (0.353,-0.353) -- (-0.353,0.353);
\end{scope}
\begin{scope}[xshift=8cm]
\draw (0,0) circle (0.5cm);
\draw (0.353,0.353) -- (-0.353,-0.353);
\draw (0.353,-0.353) -- (-0.353,0.353);
\end{scope}
\begin{scope}[xshift=12cm]
\draw (0,0) circle (0.5cm);
\draw (0.353,0.353) -- (-0.353,-0.353);
\draw (0.353,-0.353) -- (-0.353,0.353);
\end{scope}
\end{scope}
\begin{scope} [yshift=2cm,scale=0.7]
\node at (-7,0) {$\eta_1=1,\eta_2=-1$};
\node at (-1,1) {$\hat1$};
\node at (1,1) {$\tilde1$};
\node at (3,1) {$\tilde2$};
\node at (5,1) {$\hat2$};
\node (a1) at (9,1) {$\hat3$};
\node (a2) at (7,1) {$\tilde3$};
\node at (13,1) {$\tilde4$};
\node at (11,1) {$\hat4$};
\draw (0.5,0) -- (1.5,0);
\draw (2.5,0) -- (3.5,0);
\draw (4.5,0) -- (5.5,0);
\draw (6.5,0) -- (7.5,0);
\draw (8.5,0) -- (9.5,0);
\draw (10.5,0) -- (11.5,0);
\draw (2,0) circle (0.5cm);
\draw (10,0) circle (0.5cm);
\begin{scope}[xshift=0cm]
\draw (0,0) circle (0.5cm);
\draw (0,0) circle (0.4cm);
\draw (0.285,0.285) -- (-0.285,-0.285);
\draw (0.285,-0.285) -- (-0.285,0.285);
\end{scope}
\begin{scope}[xshift=4cm]
\draw (0,0) circle (0.5cm);
\draw (0,0) circle (0.4cm);
\draw (0.285,0.285) -- (-0.285,-0.285);
\draw (0.285,-0.285) -- (-0.285,0.285);
\end{scope}
\begin{scope}[xshift=6cm]
\draw (0,0) circle (0.5cm);
\draw (0,0) circle (0.4cm);
\draw (0.285,0.285) -- (-0.285,-0.285);
\draw (0.285,-0.285) -- (-0.285,0.285);\end{scope}
\begin{scope}[xshift=8cm]
\draw (0,0) circle (0.5cm);
\draw (0.353,0.353) -- (-0.353,-0.353);
\draw (0.353,-0.353) -- (-0.353,0.353);
\end{scope}
\begin{scope}[xshift=12cm]
\draw (0,0) circle (0.5cm);
\draw (0.353,0.353) -- (-0.353,-0.353);
\draw (0.353,-0.353) -- (-0.353,0.353);
\end{scope}
\end{scope}
\begin{scope} [yshift=0cm,scale=0.7]
\node at (-7,0) {$\eta_1=-1,\eta_2=-1$};
\node at (-1,1) {$\hat1$};
\node at (1,1) {$\tilde1$};
\node at (3,1) {$\tilde2$};
\node at (5,1) {$\hat2$};
\node (a1) at (7,1) {$\hat3$};
\node (a2) at (9,1) {$\tilde3$};
\node at (11,1) {$\tilde4$};
\node at (13,1) {$\hat4$};
\draw (0.5,0) -- (1.5,0);
\draw (2.5,0) -- (3.5,0);
\draw (4.5,0) -- (5.5,0);
\draw (6.5,0) -- (7.5,0);
\draw (8.5,0) -- (9.5,0);
\draw (10.5,0) -- (11.5,0);
\draw (2,0) circle (0.5cm);
\draw (6,0) circle (0.4cm);
\draw (6,0) circle (0.5cm);
\draw (10,0) circle (0.5cm);
\begin{scope}[xshift=0cm]
\draw (0,0) circle (0.5cm);
\draw (0,0) circle (0.4cm);
\draw (0.285,0.285) -- (-0.285,-0.285);
\draw (0.285,-0.285) -- (-0.285,0.285);
\end{scope}
\begin{scope}[xshift=4cm]
\draw (0,0) circle (0.5cm);
\draw (0,0) circle (0.4cm);
\draw (0.285,0.285) -- (-0.285,-0.285);
\draw (0.285,-0.285) -- (-0.285,0.285);
\end{scope}
\begin{scope}[xshift=8cm]
\draw (0,0) circle (0.5cm);
\draw (0.353,0.353) -- (-0.353,-0.353);
\draw (0.353,-0.353) -- (-0.353,0.353);
\end{scope}
\begin{scope}[xshift=12cm]
\draw (0,0) circle (0.5cm);
\draw (0.353,0.353) -- (-0.353,-0.353);
\draw (0.353,-0.353) -- (-0.353,0.353);
\end{scope}
\end{scope}
\end{tikzpicture}
\end{equation*}
The case $\eta_1=\eta_2=-1$ is the one considered in the last section.

What distinguishes all four of these choices is that the action of the automorphism on the simple roots is the same:
\EQ{
&\sigma_-(\Balpha_1)=\Balpha_3\ ,\qquad \sigma_-(\Balpha_2)=\Balpha_2\ ,\qquad\sigma_-(\Balpha_3)=\Balpha_1\ ,\\
&\sigma_-(\Balpha_4)=-\sum_{r=1}^7\Balpha_r\ ,\\
&\sigma_-(\Balpha_5)=\Balpha_7\ ,\qquad \sigma_-(\Balpha_6)=\Balpha_6\ ,\qquad\sigma_-(\Balpha_7)=\Balpha_5\ .
\label{exc}
}
The dot products of the roots with the Cartan vectors $\Bxi_j$ are
\EQ{
\Bxi_j\cdot\Balpha_r&=\big(\eta_2,0,-\eta_2,\tfrac12(2(-1)^j+\eta_2+\eta_1),-\eta_1,0,\eta_1\big)\ .
}
In particular, $(\Bxi_1-\Bxi_2)\cdot\Balpha_r=-2\delta_{r4}$, so the energy and momentum in the gauge-fixed theory is always associated to the density $\rho_4(x)$ in all of the 4 bases.

From \eqref{exc}, we find that 
\EQ{
\beta_j(x)&=\frac{(-1)^j}2\big(\eta_2(G_3(x)-G_1(x))+\eta_1(G_5(x)-G_3(x))\\ &-\frac12(2(-1)^j+\eta_1+\eta_2)G_4(x)\big)
}
The driving terms are
\EQ{
F_{1}(x)&=-\eta_2(G_4(1/x)+u_1(x)+u_2(x))\ ,\quad F_2(x)=0\ ,\\
F_{3}(x)&=\eta_2(G_4(1/x)+u_1(x)+u_2(x))\ ,\\
F_4(x)&=\eta_2(G_3(1/x)-G_1(1/x))+\eta_1(G_5(1/x)-G_7(1/x))\\ &-(\eta_1+\eta_2)G_4(1/x)+u_1(x)-u_2(x)+(\eta_1+\eta_2)(u_1(x)+u_2(x))/2\ ,\\
F_{5}(x)&=\eta_1(G_4(1/x)+u_1(x)+u_2(x))\ ,\\ F_6(x)&=0\ ,\quad
F_{7}(x)=-\eta_1(G_4(1/x)+u_1(x)+u_2(x))\ .
\label{drt3}
}

\section{Discussion}\label{s6}

In this work, we have constructed the spectral curve of the lambda string model at the classical level as a Riemann-Hilbert problem, i.e.~in terms of some densities that satisfy the integral equations \eqref{ccv2}. The densities determine a set of resolvent functions which in term determine the quasi-momentum. An outstanding problem is to present the construction of the classical solution that corresponds to a particular curve. 

Our motivation for presenting the classical curve is to provide a way to test the quantum curve which we will present in a companion paper \cite{qsc}. In fact, in anticipation of the 
quantum curve, we can already point out that the integral equations of the classical curve \eqref{ccv2} have the form of a classical limit of a set of Bethe Ansatz equations for the supergroup $\PSU(2,2|4)$. Importantly the Bethe Ansatz equations are of XXZ type rather than the XXX type familiar from the undeformed case. 

One way to write down the Bethe Ansatz equations is to start with a set of Bethe roots $\{u_{r,j}\}$, $j=1,2,\ldots, K_r$, one set for each simple root, $r=1,2,\ldots,7$. In terms these, one defines a set of ``$Q$ functions"
\EQ{
Q_r(u)=\Phi_r(u)\mathbb Q_r(u)\ ,
} 
where $\mathbb Q_r(u)$ is a Baxter function of XXZ type, familiar from an XXZ type Heisenberg spin chain,
\EQ{
\mathbb Q_r(u)=\prod_{j=1}^{K_r}\sinh\big[\pi(u-u_{r,j})/k\big]\ ,
\label{rbb}
}
The pre-factor, or dressing factor, $\Phi_r(u)$ is rather non-trivial and will be described fully in \cite{qsc}. The Bethe Ansatz equations can be written as the following conditions on the $Q$ functions
\EQ{
\prod_{s=1}^7\frac{Q_s(u+i\Balpha_r\cdot\Balpha_s/2)}{Q_s(u-i\Balpha_r\cdot\Balpha_s/2)}=(-1)^{p_{\Balpha_r}+1}\ ,
\label{aba2}
}
which are the conventional nested Bethe Ansatz equations for the Baxter functions for the supergroup $\PSU(2,2|4)$,
 \EQ{
\prod_{s=1}^7\frac{{\mathbb Q}_s(u+i\Balpha_r\cdot\Balpha_s/2)}{{\mathbb Q}_s(u-i\Balpha_r\cdot\Balpha_s/2)}=e^{-iF_r(u)}\ ,
\label{aba}
}
for each $u\in\{u_{r,j}\}$, but with unconventional driving terms on the right-hand side. These driving terms are determined by the dressing factors $\Phi_r$.

The classical limit involves taking $\kappa^2\to\infty$ and $k\to\infty$ keeping the ratio $\kappa^2/k$ fixed, i.e.~$\xi$ and $\lambda$ are fixed. In this limit, the number of Bethe roots $K_r$ corresponding to an even root scale like $k$ and become described by a continuous density
\EQ{
\sum_{j=1}^{K_r}\cdots\longrightarrow\int dx\,\rho_r(x)\,\cdots\ .
}
Just as in the ordinary string case \cite{Gromov:2013pga}, the $Q$ functions have a WKB-like limit in terms of the quasi-momentum of the classical theory. The Bethe ansatz equations can be written as integral equations involving the densities and become precisely the conditions of the classical curve \eqref{ccv2}. The first term on the left-hand side of \eqref{ccv2} arise from the classical limit $\partial_u\mathbb Q_r(u)\longrightarrow H_r(x(u))$, the resolvent in \eqref{hde}.

\section*{Acknowledgements}

DP is supported by an STFC studentship. TJH is supported by STFC grant ST/P00055X/1.
The work of JLM is supported by AEI-Spain (FPA2017-84436-P and Unidad de Excelencia Mar\'\i a de Maetzu MDM-2016-0692), by Xunta de Galicia-Conseller\'\i a de Educaci\'on (ED431C-2017/07), and by FEDER.

\appendix
\appendixpage

\section{Gauge fixed theory}\label{a4}

The question of what is the energy, i.e.~the Hamiltonian, in the lambda string is a subtle one. The first comment is that the worldsheet Hamiltonian vanishes before gauge fixing, as one expects in a theory of gravity (on the worldsheet). After gauge fixing, a non-trivial Hamiltonian is induced because the gauge-fixing procedure involves time-dependent conditions. In the conformal gauge, these time-dependent constraints are needed to fix conformal reparameterizations which concretely mean taking $\mu_+$ and $\mu_-$ to be constant. This ``physical'' Hamiltonian is the one we are interested in here. In the case of the $\text{AdS}_5{\times} S^5$ string, after gauge fixing using a version of light-cone gauge \cite{Arutyunov:2009ga}, the physical Hamiltonian is identified with a Noether charge $\Delta-J_1$ corresponding to an isometry of the spacetime. In the lambda string, the geometry has no obvious isometries, but, nevertheless, one can still construct the physical Hamiltonian that generates time translation on the worldsheet in the gauge fixed theory.  We can then refer to appendix B of \cite{Appadu:2017xku} to demonstrate that it is a Noether charge. 

Gauge fixing on the worldsheet involves a kind of light-cone gauge. The choice of gauge involves identifying a ``vacuum", or reference, configuration. For the string in flat space $X^\mu$, $\mu=0,1,\ldots,D-1$, the reference configuration corresponds to $X^+=X^0+X^{D-1}=c\tau$, with $X^-=X^0-X^{D-1}=0$ and $X^i=0$, $i=1,\ldots,D-2$. This corresponds to a point-like string moving along a null geodesic. The transverse coordinates $X^i$ are then the physical fields on the worldsheet after gauge fixing. In the undeformed model, a suitable reference configuration is the ``BMN vacuum" \cite{Berenstein:2003gb} that also corresponds to a null geodesic in the geometry  and so has the same interpretation of a point-like string orbiting the equator of the $S^5$. For this solution the Lax connection has the form
\EQ{
\LAX_\pm=\mu z^{\pm2}\Lambda\ ,
\label{yer1}
}
for constant $\mu$ to be constant. The group field in the undeformed model takes the form
\EQ{
f(\tau,\sigma)=\exp(2\mu\tau\Lambda)\ ,
}
and the corresponding solution of the associated linear system is of the form
\EQ{
\Psi(z)=\exp\big[-\mu((z^2+z^{-2})\tau+(z^2-z^{-2})\sigma)\Lambda]\ .
}
The gauge-fixed set of configurations are built on top of the reference/vacuum configuration with a gauge-fixed Lax connection \cite{Grigoriev:2007bu}
\EQ{
\LAX_+=\gamma^{-1}\partial_+\gamma+z\psi_++\mu z^2\Lambda\ ,\qquad
\LAX_-=z^{-1}\gamma^{-1}\psi_-\gamma+\mu z^{-2}\gamma^{-1}\Lambda\gamma\ .
}
In the above the $G$-valued group field $\gamma$ and $\psi_\pm$ are a pair of Grassmann fields lying in $\mf^{(1)}$ and $\mf^{(3)}$. There are residual constraints that 
$\gamma^{-1}\partial_+\gamma$, $\partial_-\gamma\,\gamma^{-1}$ and $\psi_\pm$ must lie in the image of $\text{ad}\,\Lambda$ on the algebra. The degrees of freedom in $\gamma$, so constrained, are the analogue of the transverse coordinates $X^i$ of the flat space string. 

Now we turn to the lambda model and follow the same logic. A suitable reference, or vacuum, solution in this case is obtained by taking 
\EQ{
\LAX_\pm=\pm\mu z^{\pm2}\Lambda\ ,
\label{yer2}
}
compare \eqref{yer1}. The group field of the lambda model takes the form
\EQ{
\CF(\sigma,\tau)=\exp[2\mu(\lambda^{-1}-\lambda)\tau\Lambda]\ ,
}
which is also interpreted as a point-like closed string propagating along a null geodesic in the geometry. This corresponds to a different solution of the linear system compared with the undeformed case above,
\EQ{
\Psi(z)=\exp\big[-\mu((z^2-z^{-2})\tau+(z^2+z^{-2})\sigma)\Lambda]\ .
}
Hence, the gauge-fixed Lax connection is different by an important sign
\EQ{
\LAX_+=\gamma^{-1}\partial_+\gamma+z\psi_++\mu z^2\Lambda\ ,\qquad
\LAX_-=z^{-1}\gamma^{-1}\psi_-\gamma-\mu z^{-2}\gamma^{-1}\Lambda\gamma\ .
\label{yer3}
}

Before we continue, it is interesting to look at the plane wave limit in both the sigma and lambda models. Ignoring the fermions, the plane wave limit is obtained by linearizing around the identity for the field $\gamma=e^\phi=1+\phi+\cdots$. The equation of motion for the algebra field $\phi$ is the massive equation:
\EQ{
\partial_+\partial_-\phi\mp\mu^2[\Lambda,[\Lambda,\phi]]=0\ .
}
The $\mp$ sign here corresponds to the sigma and lambda models, respectively. The mass terms for $\phi$ have the conventional sign for the undeformed case, corresponding to the fact that geodesics nearby the BMN geodesic are converging. However, in the lambda model, the mass terms have the ``wrong" sign, corresponding to the fact that geodesics nearby the vacuum geodesic are diverging. The implication is that the plane wave limit in the lambda model around our chosen vacuum \eqref{yer2} is not self consistent: geodesics diverge away and the plane wave limit of the metric breaks down. Another characteristic of the lambda model vacuum \eqref{yer2} is that the classical solutions, the magnons, defined in the limit of large $\mu$ (the Hofmann-Maldacena limit) around the vacuum solution---na\"\i vely at least---are superluminal.\footnote{This follows because the magnons are obtained from those the undeformed model by swapping over $\tau$ and $\sigma$.} The problem is solved by realizing that magnons should be defined around a different reference configuration with $\gamma=\gamma_0\neq 1$ (so lying   within the gauge-fixed subspace):
\EQ{
\LAX_+&=\mu z^2\Lambda=\mu z^2\big(\Lambda_1+\Lambda_2\big)\ ,\\
\LAX_-&=-\mu z^{-2}\gamma_0^{-1}\Lambda\gamma_0=\mu z^{-2}
\big(-\Lambda_1+\Lambda_2\big)\ .
\label{yer4}
}
So $\gamma_0$ is a constant group element that has the property $\gamma_0^{-1}\Lambda_1\gamma_0=\Lambda_1$ and $\gamma_0^{-1}\Lambda_2\gamma_0=-\Lambda_2$.\footnote{Note that it is important here that it is only the sign of the compact element $\Lambda_2$ that changes sign because there is no group element in $F$ that can change the sign of the non-compact element $\Lambda_1$.}
Notice that the Virasoro constraints are still satisfied (as they are for all solutions in the gauge-fixed subspace).

With this choice,
\EQ{
\Psi(z)=\exp\big[-\mu(z^2+z^{-2})(\sigma\Lambda_1+\tau\Lambda_2)-\mu(z^2-z^{-2})(\sigma\Lambda_2+\tau\Lambda_1)]
}
and the group field of the reference solution has the form
\EQ{
\CF(\sigma,\tau)=\exp[2\mu(\lambda^{-1}-\lambda)\tau\Lambda_1-2\mu(\lambda^{-1}+\lambda)\sigma\Lambda_2]\,.
}
It describes a string stretched in the compact part of the group, rather than a point-like string, propagating in the time direction. In this case, the magnon solutions are precisely those constructed in \cite{Appadu:2017xku}.  Ultimately the type of gauge fixing should not be physically significant, both reference solutions lie in the same gauge-fixed subspace---and so for simplicity in this work when we describe the gauge-fixed theory we will imply the choice with reference configuration \eqref{yer2}.

With these preliminaries there are a series of steps to go through in order to identify the physical Hamiltonian and momentum of the gauge-fixed theory with the choice of vacuum solution \eqref{yer2}. We start by defining the Kac-Moody currents of the lambda model formulated as a deformed gauge WZW model
\EQ{
\JJ_\pm=-\frac k{2\pi}\big(\CF ^{\mp1}\partial_\pm\CF^{\pm1} +\CF ^{\mp1}A_\pm\CF^{\pm1} -A_\mp\big)\ .
}
The equations of motion of the gauge field yield the constraints \eqref{cox} which take the form
\EQ{
\JJ_\pm+\frac k{2\pi}\big(\Omega_\mp A_\pm-A_\mp\big)=0\ .
\label{pww}
}
In the theory of constrained Hamiltonian systems, these constraints are partly first and partly second class constraints. Fortunately we will not need to describe this in detail and we will be able to avoid most of the resulting complications.

The Poisson brackets can also be formulated directly in terms of the Lax connection in much the same way as for the undeformed model itself \cite{Magro:2008dv,Vicedo:2009sn,Vicedo:2010qd}. In fact, the formalism of \cite{Vicedo:2010qd}
applies directly to the lambda theory by replacing the twist function by the $\lambda$ dependent one \eqref{twt}. In order to define this Poisson bracket consistently, one needs to write the Lax connection in the Hamiltonian formalism. The key point is to find a formulation where the Lax connection is flat in the strong sense without imposing any constraints. This involves adding additional terms that vanish on the constraint surface. This is discussed in details in \cite{Magro:2008dv,Vicedo:2009sn,Vicedo:2010qd} in the undeformed case. Mirroring this in the lambda model, the spatial component of the Lax operator can be written as
\EQ{
\LAX_1&=\frac{2\pi}k\frac{z^4-\lambda^2}{\lambda^2-\lambda^{-2}}\big[\JJ_+^{(0)}+\lambda^{-3/2}z^{-3}\JJ_+^{(1)}+\lambda^{-1}z^{-2}\JJ_+^{(2)}+\lambda^{-1/2}z^{-1}\JJ_+^{(3)}\big]\\ &
+\frac{2\pi}k\frac{z^4-\lambda^{-2}}{\lambda^2-\lambda^{-2}}\big[\JJ_-^{(0)}+\lambda^{3/2}z^{-3}\JJ_-^{(1)}+\lambda z^{-2}\JJ_-^{(2)}+\lambda^{1/2}z^{-1}\JJ_-^{(3)}\big]\ .
}
This is equal to $\LAX_1$ in \eqref{rmm} with \eqref{LaxLM} up to the gauge fixing condition $A_0^{(0)}=0$ and the constraints $\JJ^{(0)}_\pm=\mp kA^{(0)}_1/2\pi$. Importantly, $\LAX_1$ can be written in a way that involves the twist function
\EQ{
\LAX_1=-\phi (z)^{-1}\sum_{j=1}^\infty\Big[
(\lambda^{1/2}z)^j\JJ_+^{(j)}+(\lambda^{-1/2}z)^j\JJ_-^{(j)}\Big]\ ,
\label{pup}
}
where the label $j$ on $\JJ_\pm^{(j)}$ is to be understood modulo 4.

Elements of the loop algebra $\hat\mf$ can be given an alternative Lie bracket $\hat\mf_R$:
\EQ{
[X,Y]_R=\frac12[RX,Y]+\frac12[X,RY]\ ,
\label{rbr}
}
where $R=R^++R^-$ and the algebra endomophisms are $R^+\hat f=\hat f_{\geq0}$ and $R^-\hat\mf=-\hat\mf_{<0}$ so that $R^+-R^-=1$. Explicitly, for $X(z)=\sum_nX_nz^n\in\hat\mf$,
\EQ{
R^+(X(z))=\sum_{n\geq0} X_nz^n\ ,\qquad R^-(X(z))=-\sum_{n<0}X_nz^n\ .
\label{aend}
}

The natural setting for the Poisson structure of our theory, is the double loop algebra: that is elements of the loop algebra $\widehat F$ that are functions of the spatial coordinate $\sigma$. We will need a twisted inner product on this object:
\EQ{
\big\langle a,b\rangle_{\phi}=\int_{-\pi}^\pi d\sigma\,\oint\frac{dz}{2\pi iz}\phi(z)\STr\big\{a(\sigma;z)b(\sigma;z)\big\}\ ,
\label{inner}
}
where the contour integral picks up the poles of the twist function. Note that the twisted inner product involves the 1-form $du$ in \eqref{twt}.

The Poisson bracket of the theory is simply the Poisson bracket Kac-Moody algebra
\EQ{
\big\{\JJ^a_\pm(\sigma),\JJ^b_\pm(\sigma')\big\}&=f^{ab}{}_c\JJ_\pm^c(\sigma')\delta(\sigma-\sigma')\mp\frac {k}{2\pi}\eta^{ab}\delta'(\sigma-\sigma')\ ,\\
\big\{\JJ^a_+(\sigma),\JJ^b_-(\sigma')\big\}&=0\ ,
\label{km4}
}
with $\JJ_\pm^a=\STr(T^a\JJ_\pm)$, for a basis of generators $\{T^a\}$. These Poisson brackets can be written as a Kostant-Kirillov Poisson bracket of the centrally extended double loop algebra (in $e^{i\sigma}$ and $z$) with a modified Lie bracket, the ``$R$ bracket'', which is defined for functionals $\psi_1[\LAX_1]$ and $\psi_2[\LAX_1]$ as
\EQ{
\{\psi_1,\psi_2\}[\LAX_1]=\big\langle \LAX_1,[\delta\psi_1,\delta\psi_2]_R\big\rangle_{\phi }+\omega_R(\delta\psi_1,\delta\psi_2)\ .
\label{gbb}
}
Here, the functional derivatives $\delta\psi_i$ are elements of the Lie algebra $\hat\mf_R$ which are defined via
\EQ{
\big\langle \mathfrak A,\delta\psi\big\rangle_{\phi }=\frac d{dr}\psi[\LAX_1+r\mathfrak A]\Big|_{r=0}
\label{rxt}
}
for arbitrary $\mathfrak A$. The central extension takes the form
\EQ{
\omega_R(X,Y)=\frac12\big\langle R(\partial_1X),Y\rangle_{\phi }+\frac12\big\langle \partial_1X,R(Y)
\big\rangle_{\phi }\ .
}

The Hamiltonian equation of motion defined by $\psi[\LAX_1]$ with respect to \eqref{gbb} takes the form of a co-adjoint action:
\EQ{
\frac{\partial\LAX_1}{\partial t}&=-\text{ad}_R^* \delta\psi\cdot\LAX_1\ ,
\label{cup}
}
where the $R$ co-adjoint action is defined in terms of the usual adjoint action via
\EQ{
\text{ad}_R^*X\cdot Y=\frac12\text{ad}^*(RX)\cdot Y+\frac12R^*\text{ad}^*X\cdot Y\ ,
\label{gg3}
}
where $\text{ad}^*X\cdot\LAX_1=[X,\partial_1+\LAX_1]$ and $\langle R(X),Y\rangle_{\phi }=\langle X,R^*(Y)\rangle_{\phi }$ implying 
\EQ{
R^*=- \varphi ^{-1}R\varphi\ ,
}
where $\varphi(z)=\phi(z)/z$.  

The key result here, is that \eqref{cup} defines a Hamiltonian flow on the phase space. We now will identify a particular $\psi[\LAX_1]$ such that $t$ is the physical time $\tau$. This will identify the physical Hamiltonian.

We begin by performing the particular gauge transformation in the loop group of the spatial component of the Lax operator as in \eqref{gt8} such that $[\LAX^{(\infty)}_1(z),\Lambda]=0$. The gauge transformation $\Phi^{(\infty)}(z)$ can be constructed by expanding around $z=\infty$. We can equivalently expand around $z=0$ with $[\LAX^{(0)}_1(z),\Lambda]=0$. For instance, around $z=\infty$,\footnote{Implicitly we working on shell for the Virasoro constraints. The flows we define will be seen to lie within the constrained subspace of the phase space.}
\EQ{
\LAX^{(\infty)}_1(z)=\mu z^{-2}\Lambda+\sum_{n=-1}^\infty\LAX^{(\infty)}_{1,n}z^{-n}\ ,\qquad
\Phi^{(\infty)}(z)=\sum_{n=0}^\infty\Phi^{(\infty)}_nz^{-n}\ .
}
It is importantly that the gauge transformation $\Phi^{(\infty)}(z)$ and transformed Lax operator $\LAX^{(\infty)}_1(z)$ are local functions of the fields and their derivatives (e.g.~see \cite{BBT}).

It follows that
\EQ{
[\partial_1+\LAX_1,\Phi^{(\infty)}\Lambda\Phi^{(\infty)-1}]=0\ ,
\label{cow}
}
Using this fact allows us to define a flow $\eta$ associated to any function $\zeta(z)$ such that $\zeta(z)\Lambda$ is in the loop algebra, so $\zeta(iz)=-\zeta(z)$. The flow can be written in two equivalent ways
\EQ{
\frac{\partial\LAX_1}{\partial \eta}=[R^\pm(\Phi^{(\infty)} \zeta\Lambda \Phi^{(\infty)-1}),\partial_1+\LAX_1]\ .
\label{iqp}
}
The flow can be written in terms of $R=R^++R^-$ as
\EQ{
\frac{\partial\LAX_1}{\partial \eta}=\frac12[R(\Phi^{(\infty)}\zeta\Lambda\Phi^{(\infty)-1}),\partial_1+\LAX_1]\ .
\label{iqp2}
}
It follows that these flows on the Lax operator take the form of zero curvature condition
\EQ{
[\partial_\eta+\LAX_\eta,\partial_1+\LAX_1]=0\ ,\qquad\LAX_\eta=-\frac12R(\Phi^{(\infty)}\zeta\Lambda\Phi^{(\infty)-1})\ .
}
All the flows for different $\zeta(z)$ mutually commute so there is an integrable hierarchy of flows.

The final piece of this analysis is to show that the flows are Hamiltonian with respect to the Poisson bracket \eqref{gbb}. Let us define the Hamiltonian
\EQ{
H^{(\infty)}=\frac12\big\langle \zeta\Lambda,\LAX^{(\infty)}_1\big\rangle_\phi\ .
\label{HHa}
}
It follows that the functional derivative $\delta H^{(\infty)}$ defined in \eqref{rxt} is
\EQ{
\delta H^{(\infty)}=\frac12\Phi^{(\infty)}\zeta\Lambda\Phi^{(\infty)-1}\ .
} 
This implies that the flow can be written in Hamiltonian form \eqref{cup} with $\psi=H^{(\infty)}$.

The same argument applies for the expansion around $z=0$, there are commuting flows generated by Hamiltonians
\EQ{
H^{(0)}=\frac12\big\langle \zeta'\Lambda,\LAX^{(0)}_1\big\rangle_\phi\ .
\label{HHa2}
}

Let us now identified the flow associated to the worldsheet time $\tau$. We know that near $z=0$, $\LAX_0=-\mu z^{-2}\Lambda+\cdots$ and near $z=\infty$, $\LAX_0=\mu z^2\Lambda+\cdots$, therefore the physical Hamiltonian $H$ is identified as $H^{(\infty)}+H^{(0)}$ with $\zeta(z)=\mu z^2$ and $\zeta'(z)=\mu z^{-2}$. Performing the $z$ integral in \eqref{HHa}
\EQ{
H&=\frac\mu2\big\langle z^2\Lambda,\LAX^{(\infty)}_1(z)\big\rangle_\phi-\frac\mu2\big\langle z^{-2}\Lambda,\LAX^{(0)}_1(z)\big\rangle_\phi\\
&\propto\int_{-\pi}^\pi d\sigma\,\STr\Big\{\Lambda\big(\lambda\LAX_1^{(\infty)}(\lambda^{1/2})-\lambda^{-1}\LAX_1^{(\infty)}(\lambda^{-1/2})\big)\\ &\qquad\qquad-\Lambda\big(\lambda^{-1}\LAX_1^{(0)}(\lambda^{1/2})-\lambda\LAX_1^{(0)}(\lambda^{-1/2})\big)\Big\}\ .
}
This identifies the physical Hamiltonian as \eqref{suc1} in the main text. Similarly the worldsheet momentum is identified as $H^{(\infty)}-H^{(0)}$ giving \eqref{suc2} in the text.

The coda to this discussion is that the associated current $\EuScript J(z)$ has been shown in appendix B of \cite{Appadu:2017xku} to be a Noether current.

\end{document}